\documentclass{jfm}
\usepackage[utf8]{inputenc}
\usepackage{graphicx}
\usepackage{float}
\usepackage{subcaption}
\usepackage{textcomp}\usepackage{gensymb} 
\usepackage{xcolor} 
\usepackage{amsmath} 
\usepackage{amssymb} 
\usepackage[colorlinks=true,citecolor=blue,linkcolor=blue]{hyperref}
\usepackage{natbib}

\def\mean#1{\left< #1 \right>}
\def\meanR#1{\tilde{ #1 }}
\newcommand{\dif}{\mathrm{d}}
\newcommand{\ReNumber}{{Re}}
\newcommand{\PrNumber}{{Pr}}
\newcommand{\NuNumber}{{Nu}}
\newcommand{\PeNumber}{{Pe}}
\newcommand{\StNumber}{{St}}
\newcommand{\DR}{{DR}}
\newcommand{\meanDvT}{{\big<\meanR{v}\meanR{T}\big>}}

\shorttitle{Heat transfer increase by convection}
\shortauthor{J. Sundin, U. Ciri, S. Leonardi, M. Hultmark and S. Bagheri}

\title{Heat transfer increase by convection in liquid-infused surfaces for laminar and turbulent flows}

\author{Johan Sundin\aff{1}\corresp{\email{johasu@mech.kth.se}}, Umberto Ciri\aff{2}, Stefano Leonardi\aff{2}, Marcus Hultmark\aff{3} \and Shervin Bagheri\aff{1}}

\affiliation{\aff{1}FLOW Centre, Dept.~Engineering Mechanics, KTH, Stockholm SE-100 44, Sweden \aff{2}Dept.~Mechanical Engineering, The University of Texas at Dallas, Richardson, TX 75080, USA \aff{3} Mechanical and Aerospace Engineering Dept., Princeton University, Princeton, NJ 08544, USA}

\begin{document}
\maketitle

\begin{abstract}
Liquid-infused surfaces (LIS) can reduce friction drag in both laminar and turbulent flows. However, the heat transfer properties of such multi-phase surfaces have still not been investigated to a large extent. We use numerical simulations to study conjugate heat transfer of liquid-filled grooves. It is shown that heat transfer can increase for both laminar and turbulent liquid flows due to recirculation in the surface texture. For the increase to be substantial, the thermal conductivity of the solid must be similar to the thermal conductivity of the fluids, and the recirculation in the grooves must be sufficiently strong (P\'eclet number larger than 1). The ratio of the surface cavity to the system height is an upper limit of the direct contribution from the recirculation. While this ratio can be significant for laminar flows in microchannels, it is limited for turbulent flows, where the system scale (e.g.~channel height) usually is much larger than the texture height. However, heat transfer enhancement on the order of $10\%$ is observed (with a net drag reduction) in a turbulent channel flow at a friction Reynolds number $\ReNumber_\tau \approx 180$. It is shown that the turbulent convection in the bulk can be enhanced indirectly from the recirculation in the grooves.
\end{abstract}

\section{Introduction}
\begin{figure}
  \centering
  \includegraphics[width=\textwidth]{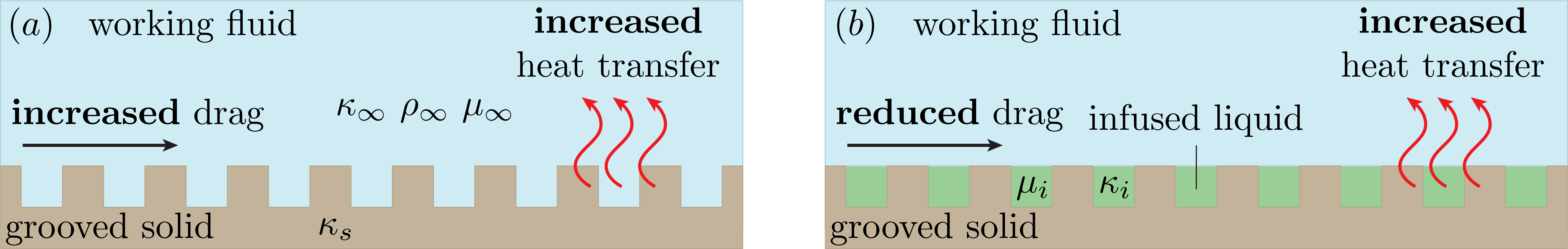}
  \caption{($a$) Current practice to increase heat transfer through wall modification and ($b$) suggested method by using LIS. The working (external) fluid and the infused liquid are shown in blue and green, respectively. The solid is depicted in brown. }
  \label{fig:sketch}
\end{figure}

The simultaneous increase of heat transfer and reduction of drag in laminar and turbulent fluid flows has turned out to be a considerable challenge. Such technology could lower the power input required for driving the flow in heat exchangers, microprocessors, and other thermal systems. 
%
%
A common technique to increase heat transfer in applications is to use rough or modified walls (fig.~\ref{fig:sketch}a). While roughness, grooves, blades, ridges, and corrugations increase the surface heat flux, they also increase the momentum transport, resulting in increased drag \citep{rohsenow98}. Enhancement of heat transfer in pipes, channels, and ducts can also be achieved by miniaturisation since this increases the surface-to-volume ratio. Effectively, more liquid is then in contact with the surface for a given volume flow rate. However, reduced dimensions also lead to increased wall friction so that the required pressure drop (and pumping power) increases. 

Both heat transfer increase and drag reduction can be achieved with complex surfaces having multiple degrees of freedom to tune the surface-flow interaction. One example is a liquid-infused surface (LIS). The infused liquid creates a dynamic interface that induces a slipping effect on the external flow while affecting wall-normal velocity fluctuations (fig.~\ref{fig:sketch}b). The careful design of texture topology, liquid viscosity, thermal conductivity, and surface tension enables the precise tuning of transport processes.

In previous studies, superhydrophobic surfaces (SHS) have been used to reduce friction while maintaining the benefits of reduced system size (see the review by \citet{gong21}). Similar effects could be accomplished with a LIS. An analytical model of heat transfer in pipes with SHS or LIS, taking the reduced friction into account, was recently developed by \citet{hatte21}. Indeed, reducing the pipe radius increased the advantage of using SHS and LIS.
%
\citet{ciri21} designed LIS and SHS that increased heat transfer and decreased drag in direct numerical simulations of a turbulent channel flow. Out of the many investigated surface configurations, this occurred only for SHS with longitudinal grooves, dynamic interfaces, and gas-liquid thermal conductivity ratio one. The solid structures were also assumed to have a constant temperature. A thermal conductivity ratio of one and isothermal textures are hard to realise in practice, however.

If the thermal conductivity of the solid texture is similar to a contiguous liquid, the heat flux in neither phase is trivial. These solid-liquid combinations are not rare in applications. For example, water has a higher thermal conductivity than polymer polydimethylsiloxane (PDMS). LIS made from such materials can be expected to have a very different heat flux compared to LIS with isothermal surface textures.
For metal solids, liquid metals may be used as infused liquids. There has recently been an increased interest in using liquid gallium alloys with low toxicity in stretchable electronics \citep{dickey17}. The thermal conductivity of gallium is comparable to that of steel. An overview of some properties of representative solids and liquids is given in tab.~\ref{tab:properties}. 


Using liquid and solid properties similar to those of existing materials, we have investigated whether convection in the texture of LIS with transverse grooves can increase the surface heat flux, considering both laminar and turbulent flows. This mode of heat transfer has either been neglected or has not been evaluated closer in earlier works. The study was performed by numerically solving the equations for the velocity and temperature fields. We considered the heat transfer between two external boundaries of different temperatures. The heat transport mechanisms of this setup are shown in fig.~\ref{fig:problemDefinition}a. 

\begin{table}
\centering
\begin{tabular}{lllllllllll}
Material                    & $\kappa$ [W/(mK)]  & $\mu$ [mPas] & $\rho$ [kg/m$^3$] & $c_p$ [kJ/(kgK)] \vspace{0.2cm} \\ 
Water                       & 0.60 (R)           & 1.0 (R)      & 1000 (R)           & 4.18 (R)       \\
Polydimethylsiloxane (PDMS) & 0.16 (M)           & --           & 970 (M)            & 1.5 (M)             
\vspace{0.1cm} \\
Gallium (liquid)            & 29 (L)             & 1.97 (L)        & 6080 (L)           & 0.35 (L)        \\
Steel                       & 45 (R)             & --            & 7800 (R)                     & 0.43 (C)              
\vspace{0.1cm} \\
Hexane                      & 0.120 (R)         & 0.33 (VB)    & 655 (VB)          & 2.23 (R)        \\
Heptane                     & 0.124 (R)         & 0.43 (VB)    & 684 (VB)          & 2.20 (R)        \\
Dodecane                    & 0.135 (K1)          & 1.4 (VB)     & 750 (VB)          & 2.19 (K2)      \\
\end{tabular}
\caption{Overview of thermal conductivity, $\kappa$, viscosity, $\mu$, density, $\rho$, and isobaric specific heat capacity, $c_p$,  for some fluids and solids. (R, \citet{rohsenow98}; M, \citet{mark09}; L, \citet{liu2019}; C, \citet{cardarelli18}; VB, \citet{buren17}; K1, \citet{kashiwagi82}; K2, \citet{khasanshin03})} 
\label{tab:properties}
\end{table}

The remainder of the article has the following structure. Governing equations and definitions of relevant quantities are presented in sec.~\ref{sec:governingEquations}. In sec.~\ref{sec:varyingSolidConductivity}, it is shown that the contribution from the convection in the texture to the total heat flux is necessary to consider when the thermal conductivity of the solid structures is similar to the infused liquid.
Setting the solid and liquid thermal conductivity to the same value, the relation between the surface convection, liquid, flow and geometric properties is described in sec.~\ref{sec:propertiesDispersiveConvection}. 
Results from turbulent flow simulations are presented in sec.~\ref{sec:turbulence}. 
Conclusions are given in sec.~\ref{sec:conclusions}.

\section{Governing equations}
\label{sec:governingEquations}
We consider the momentum, continuity, and energy equations for an incompressible system,
\begin{gather}
      \rho \frac{\partial \mathbf{u}}{\partial t} + \rho\left (\mathbf{u} \cdot \nabla\right ) \mathbf{u} = -\nabla P +  \nabla \cdot \mu\left(\nabla \mathbf{u} + (\nabla \mathbf{u})^\mathrm{T}\right) = 0,\label{eq:dim_momentum_equation}\\
      \nabla \cdot \mathbf{u} = 0, \label{eq:dim_continuity}\\
   \rho c_p \frac{\partial T}{\partial t} + \rho c_p\mathbf{u} \cdot \nabla T = \nabla \cdot \kappa\nabla T,
    \label{eq:dim_heat_equation}
\end{gather}
where $\rho$ is the density, $\mathbf{u}$ is the fluid velocity, $P$ is the pressure, $\mu$ is the fluid viscosity, $c_p$ is the specific heat capacity, $T$ is the temperature,, and $\kappa$ is the thermal conductivity. The streamwise coordinate is $x$, the wall-normal $y$, with $y = 0$ at the surface, and the spanwise $z$. 
Eqs.~\eqref{eq:dim_momentum_equation} and \eqref{eq:dim_continuity} are valid in the domains occupied by the external fluid and the infused (internal) liquid. In the region occupied by the solid, only eq.~\eqref{eq:dim_heat_equation} is valid and $\mathbf{u} = 0$.

The differences between the fluid properties can be expressed through the viscosity ratio $\mu_i/\mu_\infty$, the density ratio $\rho_i/\rho_\infty$, the specific heat capacity ratio $c_{p,i}/c_{p,\infty}$, and the thermal conductivity ratio $\kappa_i/\kappa_\infty$, where the subscripts $i$ and $\infty$ refer to the infused and external fluids, respectively.
For example, \citet{rosenberg16} and \citet{buren17} used a heptane-water system (among others) 
with $\mu_i/\mu_\infty = 0.43$, $\kappa_i/\kappa_\infty = 0.21$, and $c_{p,i}/c_{p,\infty} = 0.53$ (see also tab.~\ref{tab:properties}). 
%
For simplicity and clarity, we have assumed that densities and specific heat capacities in fluids and solid, and thermal conductivities in the fluids, are equal, i.e.~
\[\rho = \rho_s = \rho_i = \rho_\infty, \quad c_p = c_{p,s} = c_{p,i} = c_{p,\infty}, \quad \textrm{and}\quad  \kappa_i = \kappa_\infty,\]
where a subscript $s$ denotes solid properties. 

%

The velocity satisfies the no-slip and impermeability conditions at solid boundaries. Across liquid-liquid interfaces, the velocity and the shear stress are continuous. We have neglected interface deformation, which is equivalent to imposing an infinite surface tension. Consequently, the wall-normal velocity component $v$ is zero at interfaces. The temperature and the (instantaneous and local) heat flux are continuous at solid boundaries and interfaces. Also, the temperature is a passive scalar, i.e.~the momentum equation is independent of the energy equation.  

\begin{figure}
  \centering
  \includegraphics[height=5.0cm]{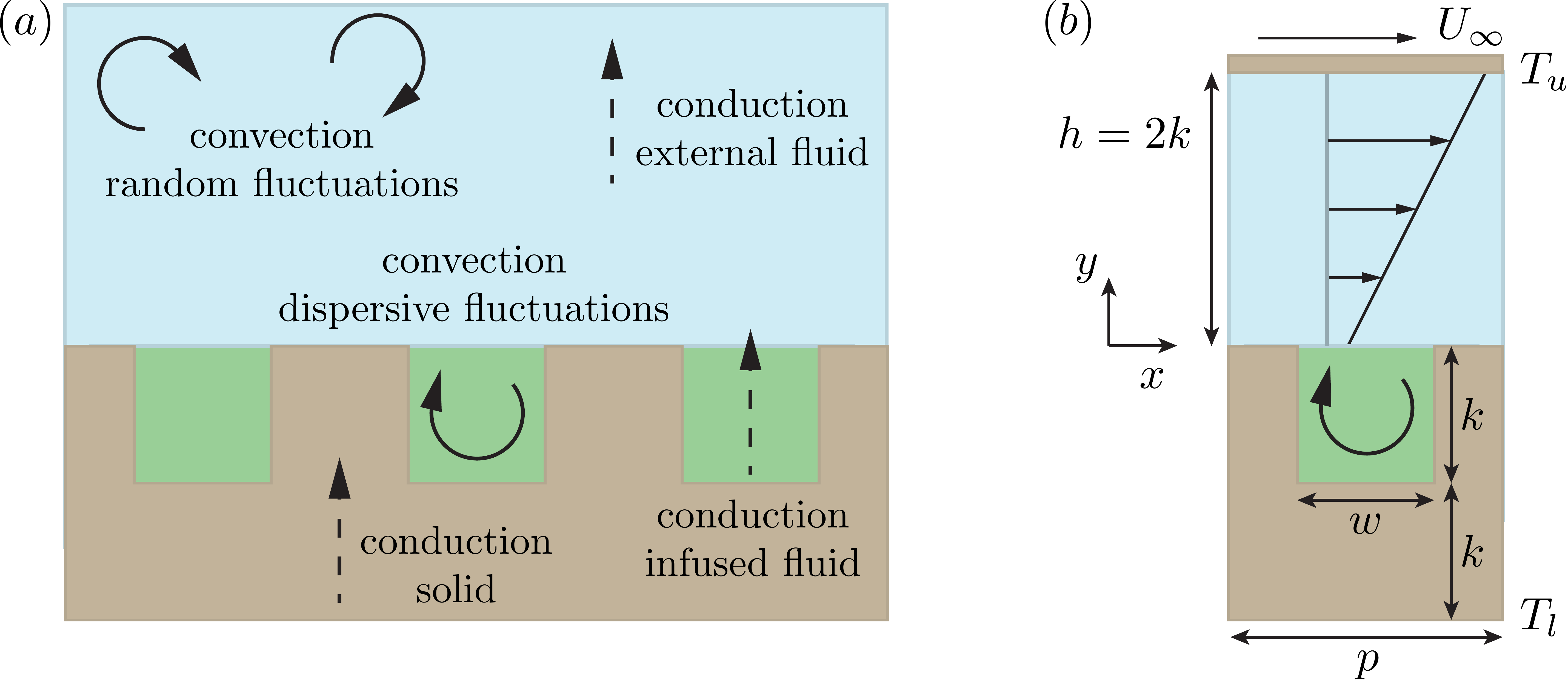}

  \caption{($a$) Heat transport mechanisms contributing  to the heat flux. ($b$) Schematic illustration of an interface unit cell for LIS. The colours of the fluids and the solid are the same as in fig.~\ref{fig:sketch}. }
  \label{fig:problemDefinition}
\end{figure}

The surface-averaged heat flux, $q$, can be decomposed into different contributions using the Fukagata, Iwamoto, and Kasagi (FIK) identity of the energy equation \citep{fukagata05}. In this decomposition, we consider a channel of height $h$ and grooves of depth $k$ with a solid slab of the same thickness beneath (fig.~\ref{fig:problemDefinition}b). The derivation is outlined in appendix \ref{sec:fikIdentities}. 
The second term of eq.~\eqref{eq:dim_heat_equation} gives rise to two convection terms, and the third term gives rise to three conduction terms. In total, the expression for $q$ is
\begin{multline}
    q = - \frac{\kappa_\infty}{h + 2k}\int_{0}^h\mean{\frac{\partial T}{\partial y}}\dif y - \frac{\kappa_s}{h + 2k}\int_{-2k}^0\mean{\frac{\partial T}{\partial y}\chi_s}\dif y \\ - \frac{\kappa_i}{h + 2k}\int_{-2k}^0\mean{\frac{\partial T}{\partial y}(1 - \chi_s)}\dif y +  \frac{\rho c_p}{h + 2k}\int_{-k}^h\mean{v'T'}\dif y + \frac{\rho c_p}{h + 2k}\int_{-k}^h\meanDvT\dif y \\ = q_{\mathrm{cond},\infty} + q_{\mathrm{cond},s} + q_{\mathrm{cond},i} + q_{\mathrm{conv},r} + q_{\mathrm{conv},d},
    \label{eq:qDecomposition}
\end{multline}
where $\chi_s$ is an indicator function equal to $1$ in the solid and $0$ elsewhere. The operator $\mean{}$ denotes the average in time and the streamwise and spanwise directions. We have separated the convective flux $\mean{vT}$ into a dispersive and a random component, $\meanDvT$ and $\mean{v'T'}$, respectively (appendix \ref{sec:flowDecomp}). 

The first three terms on the last line of eq.~\eqref{eq:qDecomposition} describe conduction in the external liquid, the solid, and the infused liquid, respectively. The following two terms correspond to convection from random (turbulent) fluctuations and dispersive (or roughness coherent) fluctuations in the vicinity of the surface, respectively. Each of these terms has its counterpart before the equality sign in the same order. They also correspond to the heat transport mechanisms represented in fig.~\ref{fig:problemDefinition}a. It is mainly the recirculation in the grooves that gives rise to $\meanDvT$, which results in $q_{\mathrm{conv},d}$. 
%
At every wall-normal location, $q$ needs to be the same since there is no heat source inside the fluids or the solid. 

The steady laminar flow problem was solved in an interface unit cell, 
illustrated in fig.~\ref{fig:problemDefinition}b. 
These simulations were performed using the finite element solver FreeFem++ \citep{hect12, lacis20}. The height of the channel was $h = 2k$ so that the total domain height was $4k$. At the upper boundary, we imposed constant shear stress and no normal stress. 
In the streamwise direction, periodic boundary conditions were imposed. We applied constant temperatures $T_u$ and $T_l$ at the upper and the lower boundaries, respectively.
More details about the simulation setup can be found in appendix \ref{sec:numericalMethodsFreeFem}. 
Details about the turbulent simulations are given in sec.~\ref{sec:turbulence}. A Reynolds number and a P\'eclet number based on the external flow quantities were defined as $\ReNumber_\infty = \rho U_\infty h/\mu_\infty$ and $\PeNumber_\infty = \rho c_p U_\infty h/\kappa_\infty$, respectively, where $U_\infty$ is the streamwise velocity at the top boundary. The corresponding Prandtl number is $\PrNumber_\infty = \PeNumber_\infty/\ReNumber_\infty=c_p\mu_\infty/\kappa_\infty$. 

\section{Heat flux for varying solid conductivity}
\label{sec:varyingSolidConductivity}
In laminar flow, there are no random fluctuations that transport heat. It follows that in the decomposition of the heat flux, $q_{\mathrm{conv},r} = 0$. Eq.~\eqref{eq:qDecomposition} reduces to
\begin{equation}
    q = q_{\mathrm{cond},\infty} + q_{\mathrm{cond},s} + q_{\mathrm{cond},i} + q_{\mathrm{conv},d}.
    \label{eq:qVaryingSolidDecomposition}
\end{equation}
We will use the smooth-wall heat flux as a reference for the LIS simulations. If the surface of the laminar flow is smooth, two additional simplifications can be made. Since there is no spatially varying texture, the dispersive convection is zero. Neither is there infused liquid conducting heat. The heat flux only depends on the conduction in the solid and the external fluid and can be evaluated to (appendix \ref{sec:fikIdentitiesLaminar}) 
\begin{equation}
    q^0 = \frac{T_l - T_u}{2k/\kappa_s + h/\kappa_\infty}.
    \label{eq:q0}
\end{equation}
Notice that $q^0$ also depends on $\kappa_s$, meaning that the reference heat flux changes with $\kappa_s$.

\begin{figure}
  \centering
  \makebox[\textwidth] {
  \begin{subfigure}{0.48\textwidth}
      \includegraphics[height=4.5cm]{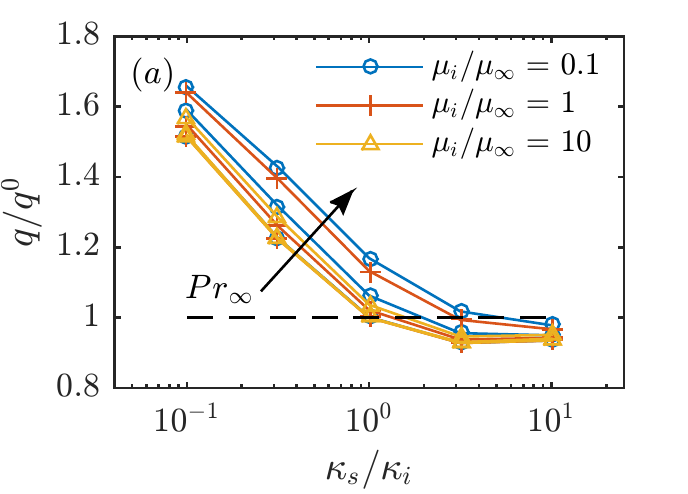}
      \captionlistentry{}
      \label{fig:NuKappa}
  \end{subfigure}
  \begin{subfigure}{0.48\textwidth}
      \includegraphics[height=4.5cm]{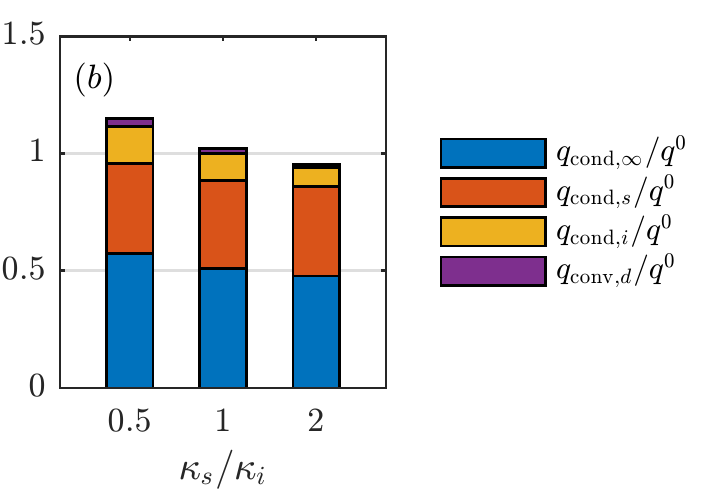}
      \captionlistentry{}
      \label{fig:qBudget}
  \end{subfigure}
  }
  \caption{($a$) The change in $q/q^0$ because of varying $\kappa_s$ for three different viscosity ratios and $\PrNumber_\infty = 1$, $10$, and $100$ at $\ReNumber_\infty = 100$. Each symbol represents a simulation result, and the lines between them are to guide the eye. The dashed line shows $q/q^0 = 1$. The ratio $q/q^0$ increases with increasing $\PrNumber_\infty$ for a fixed $\kappa_s/\kappa_i$ and $\mu_i/\mu_\infty$. The curves for $\PrNumber_\infty = 1$ are almost on top of each other.
  ($b$) The different contributions to $q$ for  $\kappa_s/\kappa_i = 0.5$, $1$, and $2$ at $\ReNumber_\infty = 100$,  $\mu_i/\mu_\infty = 1$, and $\PrNumber_\infty = 10$.
  }
  \label{fig:varyingKappaS}
\end{figure}

In fig.~\ref{fig:NuKappa}, $q/q^0$ is shown for varying solid conductivity ratios, $\kappa_s/\kappa_i$. The applied shear stress corresponds to $\ReNumber_\infty = 100$ (neglecting the slight increase in $U_\infty$ due to the finite slip velocity). We consider three values of the Prandtl number, $\PrNumber_\infty = 1$, $10$, and $100$, and three viscosity ratios, $\mu_i/\mu_\infty = 0.1$, $1$, and $10$. For comparison, if the external fluid is water, $\PrNumber_\infty = 7$, and the range of viscosity ratios covers all liquids of tab.~\ref{tab:properties}. The pitch (groove center-to-center distance) was $p = 2k$, and the groove width $w = k$. 

When $\kappa_s/\kappa_i = 1$, there is an increase in $q/q^0$ from unity because of convection. For smaller $\kappa_s/\kappa_i$ -- when the solid is a poor heat conductor compared to the infused liquid -- the average thermal conductivity of the composite surface is higher than for the solid alone, resulting in an even more pronounced increase in $q/q^0$. However, when $\kappa_s/\kappa_i$ is increased above unity, there is eventually a decrease in the surface heat flux. At about $\kappa_s/\kappa_i = 3$, the simulations result in $q/q^0 < 1$, except for the highest $\PrNumber_\infty$ and the smallest viscosity ratio. It is thus necessary that $\kappa_s \lesssim \kappa_i$ to have an increase in the heat flux.

\begin{figure}
  \centering
  \begin{subfigure}{0.31\textwidth}
      \includegraphics[height=7.5cm]{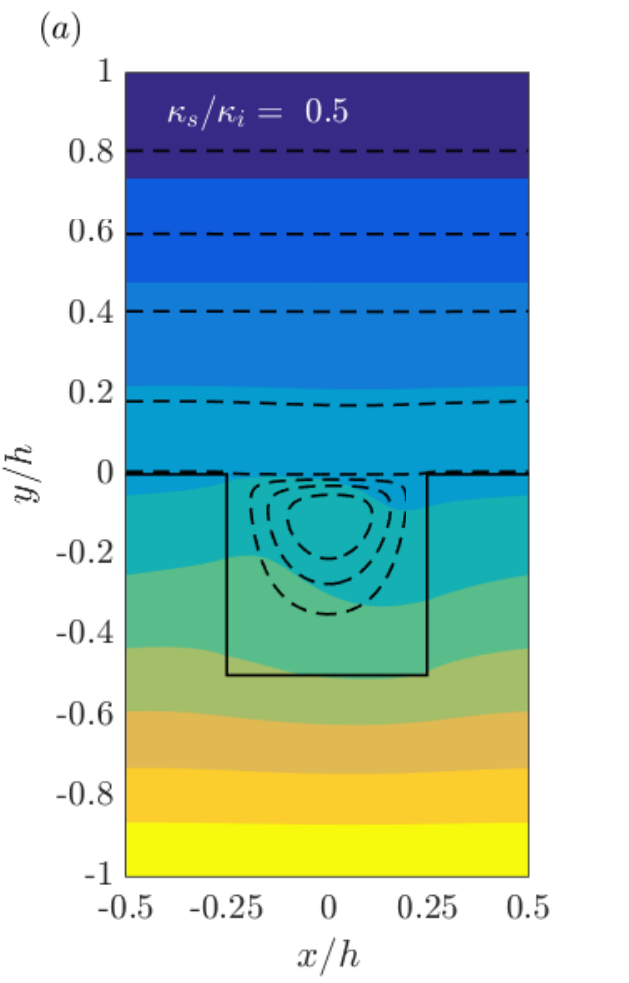}
      \captionlistentry{}
      \label{fig:tempStreamlines1}
  \end{subfigure}
    \begin{subfigure}{0.27\textwidth}
      \includegraphics[height=7.5cm]{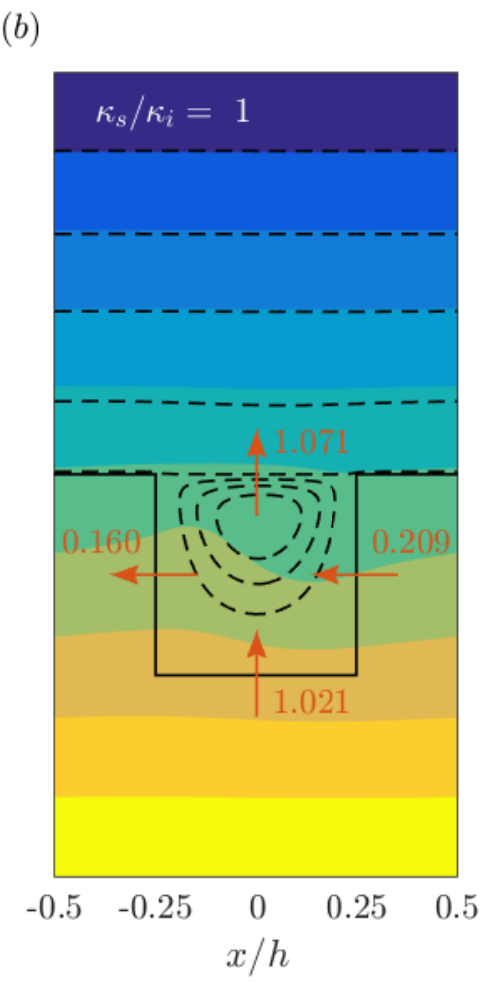}
      \captionlistentry{}
      \label{fig:tempStreamlines2}
  \end{subfigure}
  \begin{subfigure}{0.34\textwidth}
      \includegraphics[height=7.5cm]{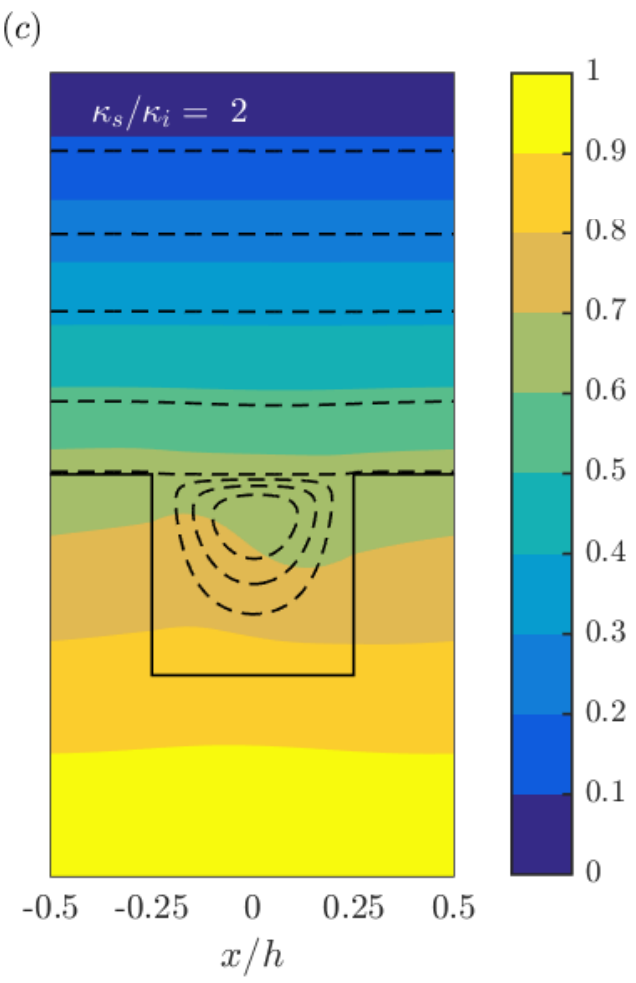}
      \captionlistentry{}
      \label{fig:tempStreamlines3}
  \end{subfigure}
  \caption{Filled contour plots of the temperature at $\ReNumber_\infty = 100$, $\mu_i/\mu_\infty = 1$, and $\PrNumber_\infty = 10$, for ($a$) $\kappa_s/\kappa_i = 0.5$, ($b$) $\kappa_s/\kappa_i = 1$, and ($c$) $\kappa_s/\kappa_i = 2$. The colour bar gives the corresponding temperatures, ranging from $T_u$ ($0$) to $T_l$ ($1$). Streamlines (dashed lines) indicate the velocity field, computed as contour lines of the stream function (constant increment in the groove; showing a few in the external flow). The edges of the solid are marked with solid lines. In and around the groove, there is a distortion of the temperature field due to convection. The convection increases heat transfer in all three cases. However, in ($a$), the higher thermal conductivity in the infused liquid also increases $q/q^0$. In ($c$), the solid has a higher thermal conductivity
  , reducing the benefit of the groove. The heat fluxes in and out of the groove are illustrated with arrows in ($b$). The attached numbers are the heat fluxes normalised by $q$.}
   \label{fig:tempStreamlines}
\end{figure}

The different terms of eq.~\eqref{eq:qVaryingSolidDecomposition} are illustrated in fig.~\ref{fig:qBudget} for $\PrNumber_\infty = 10$, $\mu_i/\mu_\infty = 1$, and $\kappa_s/\kappa_i = 0.5$, $1$, and $2$. In the case of the poor solid conductor ($\kappa_s/\kappa_i = 0.5$), $q/q^0$ would be greater than unity even without convection, because of the heat conduction in the infused liquid ($q_{\textrm{cond},i}$).  
For a good solid conductor ($\kappa_s/\kappa_i = 2$), the convection ($q_{\textrm{cond},d}$) cannot compensate for the cutout of the solid that is the groove.  
However, for $\kappa_s/\kappa_i = 1$, it is the convection that increases $q/q^0$ above unity. Contour plots of the temperature fields of these three cases are shown in fig.~\ref{fig:tempStreamlines}, together with streamlines. The streamlines indicate similarity between LIS and $d$-type roughness, with the external flow prevented from penetrating the texture \citep{jimenez04}. This behaviour is enforced by the interface and therefore also holds for larger values of $w/k$. The contour levels of the temperature field have less spacing in the region of higher conductivity. They are also distorted in and around the grooves due to convection. The convection increases the temperature on the left side of the cavity and decreases it on the right side.

We computed the heat fluxes through the different groove walls and the interface for the simulation in fig.~\ref{fig:tempStreamlines2}. These fluxes are also illustrated with arrows in the figure. 
Heat is transferred into (out of) the groove through the bottom wall (interface) and the right (left) wall. Through the interface, the heat flux was $q_I/q = 1.071$, i.e.~slightly more than the average heat flux of the surface. The heat fluxes  through the left, right and bottom walls were $q_L/q = 0.160$, $q_R/q = 0.209$, and $q_B/q = 1.021$, respectively (imbalance $|q_\mathrm{err}|/q = 1.6\cdot10^{-3}$). The increase of heat flux through the bottom wall, $q_B > q$, and sides, $q_R, q_L > 0$ is due to the convection of the cavity vortex. This convection is quantified by $q_{\mathrm{conv},d}$ and provides a net positive contribution to the total heat transfer of the surface (fig.~\ref{fig:qBudget}).


\section{Properties of the dispersive convection}
\label{sec:propertiesDispersiveConvection}
In the previous section, it was illustrated how the heat flux depends on $\kappa_s/\kappa_i$. We now focus on the dependency on the fluid and flow properties expressed through $\PrNumber_\infty$ and $\ReNumber_\infty$. Since $q_{\mathrm{conv},d}$ is responsible for $q/q^0$ increasing above unity when the thermal conductivities are similar, the discussion is limited to $\kappa_s/\kappa_i = 1$. The heat then diffuses at the same rate in the solid as in the liquid, and the sum of the conduction terms can be simplified (appendix \ref{sec:fikIdentitiesHomogeneousConductivity}), leading to
\begin{equation}
    q = q_{\mathrm{cond},\infty} + q_{\mathrm{cond},s} + q_{\mathrm{cond},i} + q_{\mathrm{conv},d} = \underbrace{\frac{\kappa_\infty}{h + 2k}(T_l - T_u)}_{q_{\mathrm{cond},\infty} + q_{\mathrm{cond},s} + q_{\mathrm{cond},i}} + \underbrace{\frac{\rho c_p}{h + 2k}\int_{-2k}^{h}\meanDvT\dif y}_{q_{\mathrm{conv},d}}.
\end{equation}
The sum of the conduction terms is denoted by $q_\mathrm{cond}$ and corresponds to $q^0$ (eq.~\ref{eq:q0}). 

\begin{figure}
  \centering
  \begin{subfigure}{0.36\textwidth}
      \includegraphics[height=4.2cm]{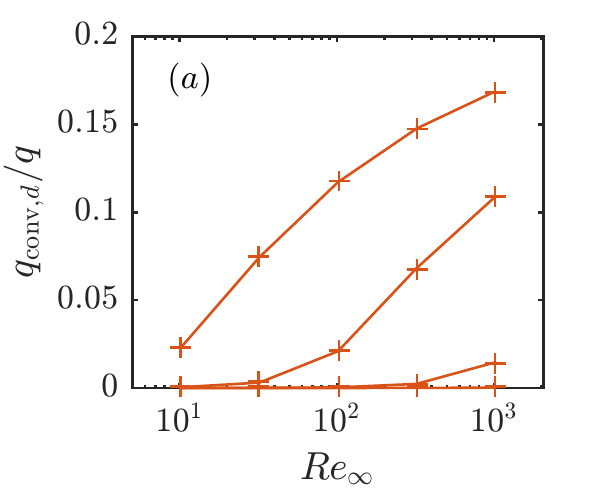}
      \captionlistentry{}
      \label{fig:NuRe}
  \end{subfigure}
  \begin{subfigure}{0.31\textwidth}
    \includegraphics[height=4.2cm]{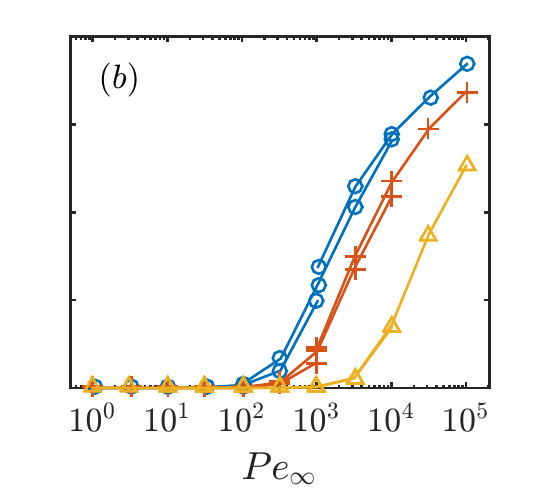}
    \captionlistentry{}
    \label{fig:NuPe}
  \end{subfigure}
  \begin{subfigure}{0.31\textwidth}
    \includegraphics[height=4.2cm]{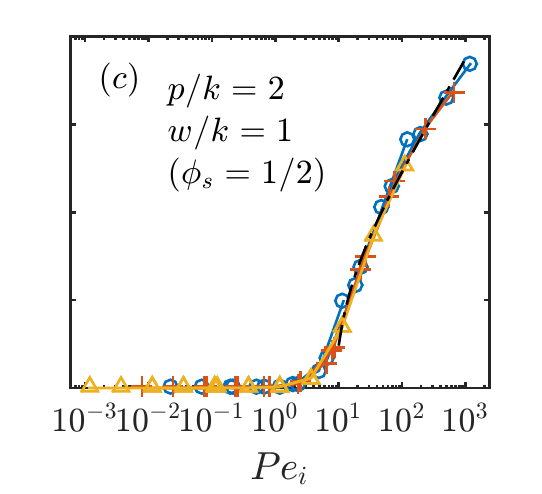}
    \captionlistentry{}
    \label{fig:NuPeI}
  \end{subfigure}
  \caption{The relative contribution to the heat flux from convection as a function of ($a$) $\ReNumber_\infty$, ($b$) $\PeNumber_\infty$, and ($c$) $\PeNumber_i$. In ($a$), results for $\mu_i/\mu_\infty = 1$ are shown; ($b$) and ($c$) contain results for $\mu_i/\mu_\infty = 0.1$, $1$, and $10$ (colours and symbols as in fig.~\ref{fig:NuKappa}). Prandtl numbers are $\PrNumber_\infty = 0.1$, $1$, $10$, and $100$. The ratio $q_{\mathrm{conv},d}/q$ increases with $\PrNumber_\infty$ for a specific $\ReNumber_\infty$. For a certain viscosity ratio, the curves collapse if expressed as a function of $\PeNumber_\infty$. However, if defined as a function of $\PeNumber_i$, they all collapse. The black dash-dotted line in ($c$) is eq.~\eqref{eq:fit}. }
  \label{fig:Pe}
\end{figure}

The relative contribution from convection to the total heat flux, $q_{\mathrm{conv},d}/q$, is shown in fig.~\ref{fig:NuRe} as a function of $\ReNumber_\infty$. These results were obtained for $\mu_i/\mu_\infty = 1$ and $\PrNumber_\infty = 0.1$, $1$, $10$, and $100$, each curve corresponding to a specific Prandtl number. For a constant $\ReNumber_\infty$, $q_{\mathrm{conv},d}/q$ increases with $\PrNumber_\infty$, as is indicated in fig.~\ref{fig:NuKappa}. If $q_{\mathrm{conv},d}/q$ instead is expressed as a function of $\PeNumber_\infty = \ReNumber_\infty\PrNumber_\infty$, the curves collapse, as shown in fig.~\ref{fig:NuPe}. 
In this figure, we also include the results for $\mu_i/\mu_\infty = 0.1$ and $\mu_i/\mu_\infty = 10$, however. For a specific $\PeNumber_\infty$, there is a dependency of $q_{\mathrm{conv},d}/q$ on the viscosity ratio. Even if $\ReNumber_\infty$ is constant, the magnitude of the flow inside the groove changes with viscosity ratio, and thereby the dispersive convection. Therefore, we need to define Reynolds and Péclet numbers that better characterize the flow at the surface.

\subsection{Surface Reynolds and Péclet numbers}
A representative velocity of the flow in the grooves is the mean velocity at the interface. If it is averaged over the whole surface, it equals the slip velocity $U_s$, which is $\mean{u}$ at $y = 0$, where $u$ is the streamwise velocity component. Using $U_s$ and $k$ as velocity and length scales, respectively, we define Reynolds and P\'eclet numbers as
\begin{equation}
    \ReNumber_i = \frac{\rho U_s k}{\mu_i} \quad \text{ and } \quad \PeNumber_i = \frac{\rho c_p U_s k}{\kappa_i},
\end{equation} 
together with a Prandtl number $\PrNumber_i = \PeNumber_i/\ReNumber_i=c_p\mu_i/\kappa_i$. Curves for different viscosity ratios collapse if $q_{\mathrm{conv},d}/q$ is expressed as a function of $\PeNumber_i$, as shown in fig.~\ref{fig:NuPeI}.

For $\PeNumber_i \gtrsim 1$, there is a noticeable increase in $q_{\mathrm{conv},d}/q$. At $\PeNumber_i = 10$, $q_{\mathrm{conv},d}/q$ is greater than 1\%. The simulation results in fig.~\ref{fig:NuPeI} are well approximated by a logarithmic function for $10^1 < \PeNumber_i < 10^3$. This set of $\PeNumber_i$ is the vital interval in practice, as it results in significant increases in heat transfer while still being attainable. The logarithmic function shown in the figure is
\begin{equation}
    \frac{q_{\mathrm{conv},d}}{q} = \frac{k}{h + 2k}0.11\ln(\PeNumber_i - 7.6),
    \label{eq:fit}
\end{equation}
found by fitting the data. This relationship is different from the power-laws used for the similar problem of a lid-driven cavity \citep{moallemi92}. 
Here, the logarithmic expression gives a better fit. For comparison, the root-mean-squared error of the logarithmic expression was 0.0054, whereas, for a power-law fit of $\alpha (\PeNumber_i^\beta - 1)$, it was 0.011, where $\alpha = 0.73k/(h + 2k)$ and $\beta = 0.11$ (with $-1$ added assuming $q_{\mathrm{conv},d} = 0$ when $\PeNumber_i = 1$). 

\begin{figure}
  \centering
  \makebox[\textwidth] {
    \begin{subfigure}{0.45\textwidth}
      \includegraphics[width=5.5cm]{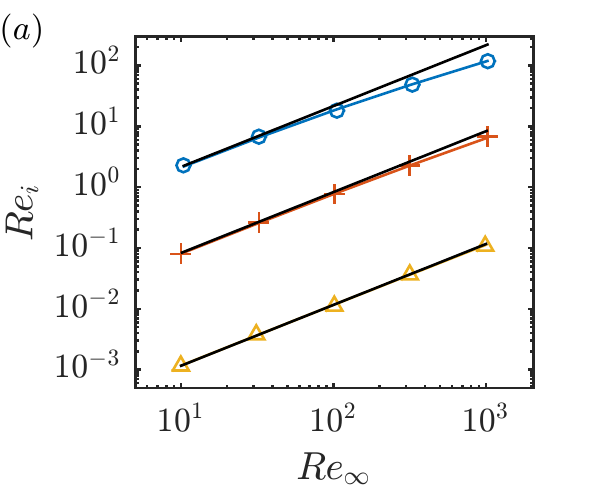}
        \captionlistentry{}
        \label{fig:re_relation}
  \end{subfigure}
  \begin{subfigure}{0.45\textwidth}
      \includegraphics[width=5.5cm]{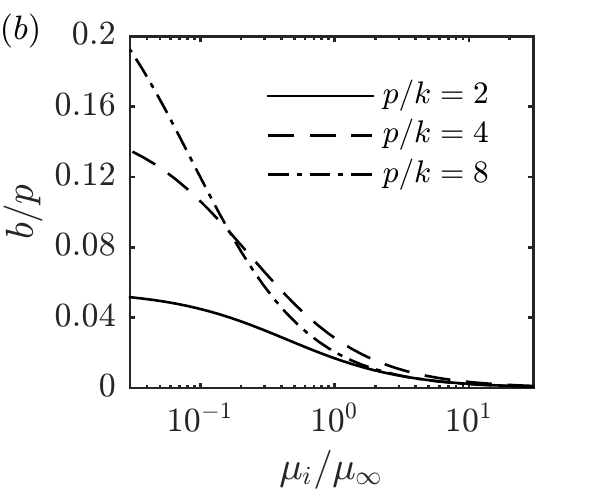}
        \captionlistentry{}
        \label{fig:slip_lengths}
  \end{subfigure}
  }
  \caption{($a$) The Reynolds number of the flow in the grooves, $\ReNumber_i$, expressed as a function of the external flow Reynolds number, $\ReNumber_\infty$, for $p/k = 2$ and three different viscosity ratios ($\mu_i/\mu_\infty = 0.1$, $1$, and $10$, with colours and symbols as in fig.~\ref{fig:NuKappa}). The relation is approximately linear for low Reynolds numbers. The linearity is made visible by the lines $\ReNumber_i = \ReNumber_\infty b/(h + b) \cdot k\mu_\infty/(h\mu_i)$, with $b$ predicted for Stokes flow (black). ($b$) The derived Stokes limit slip lengths over the pitch, $b/p$, for $p/k = 2$, $4$, and 8, and groove width $w = p - k$. }
  \label{fig:re_relation_slip_lengths}
\end{figure}

The relationships between $\ReNumber_\infty$ and $\ReNumber_i$ obtained from the simulations are plotted in fig.~\ref{fig:re_relation}. The slip velocity is related to the wall-normal derivative of $U = \mean{u}$ as $U_s = b \left.\dif U/\dif y\right|_{y = 0}$, where $b$ is the slip length, and $\left.\dif U/\dif y\right.$ is evaluated at the interface in the external fluid. Since $U_\infty = (h + b)\left.\dif U/\dif y\right|_{y = 0}$, the $\ReNumber_\infty$-to-$\ReNumber_i$ relation is a function of $b$ (appendix \ref{sec:analyticalSlipLength}). Slip lengths have also been derived analytically in the Stokes limit ($\ReNumber_\infty \rightarrow 0$) by \cite{schonecker14}, with the corresponding $\ReNumber_\infty$-to-$\ReNumber_i$ relations also shown in fig.~\ref{fig:re_relation}. There is reasonable agreement between the Stokes flow results and the simulations up to moderate $\ReNumber_i$ ($16\%$ difference for $\ReNumber_i = 19$ with $\mu_i/\mu_\infty = 0.1$). The analytical slip lengths are plotted in fig.~\ref{fig:slip_lengths} for three different pitches.

\begin{figure}
  \centering
  \begin{subfigure}{0.36\textwidth}
      \includegraphics[height=3.8cm]{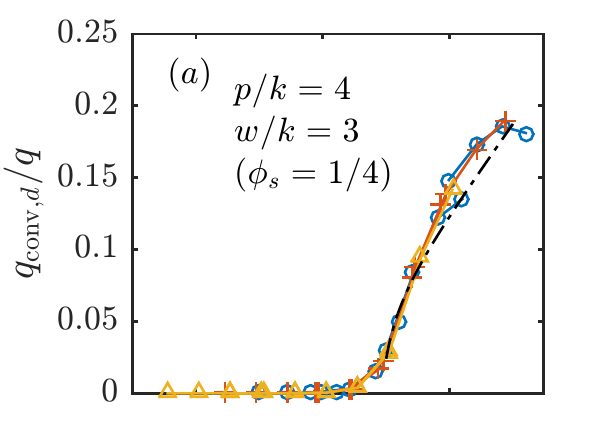}
      \captionlistentry{}
      \label{fig:NuPeOtherGeometries1}
  \end{subfigure}
  \begin{subfigure}{0.31\textwidth}
    \includegraphics[height=3.8cm]{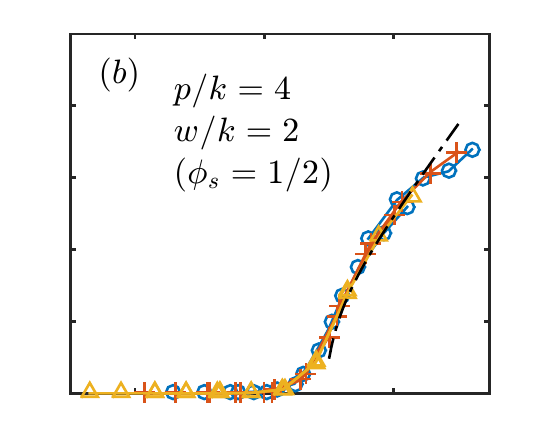}
    \captionlistentry{}
    \label{fig:NuPeOtherGeometries2}
  \end{subfigure}
  \begin{subfigure}{0.31\textwidth}
    \includegraphics[height=3.8cm]{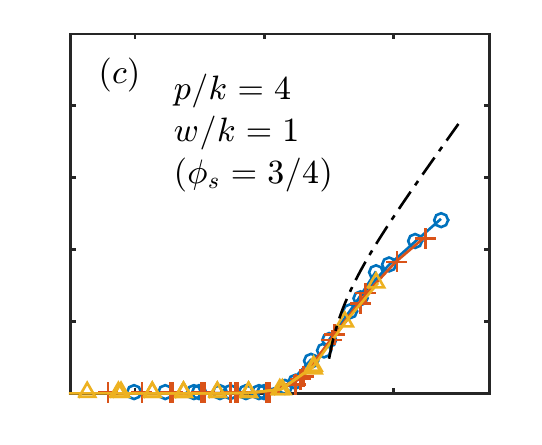}
    \captionlistentry{}
    \label{fig:NuPeOtherGeometries3}
  \end{subfigure}
    \begin{subfigure}{0.36\textwidth}
    \includegraphics[height=4.2cm]{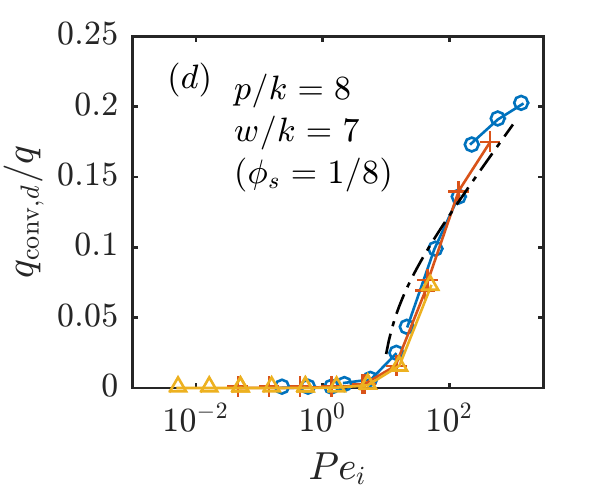}
    \captionlistentry{}
    \label{fig:NuPeOtherGeometries4}
  \end{subfigure}
  \begin{subfigure}{0.31\textwidth}
    \includegraphics[height=4.2cm]{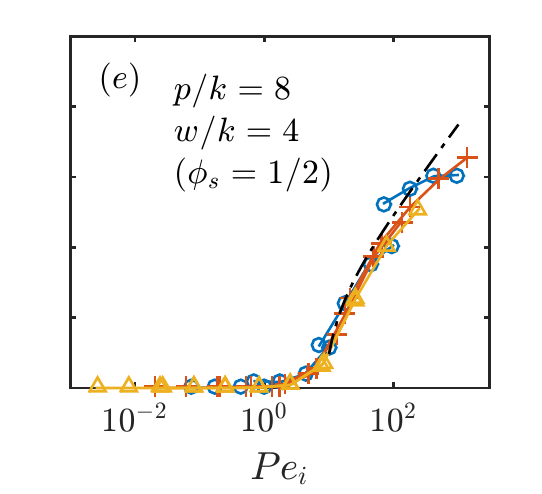}
    \captionlistentry{}
    \label{fig:NuPeOtherGeometries5}
  \end{subfigure}
  \begin{subfigure}{0.31\textwidth}
    \includegraphics[height=4.2cm]{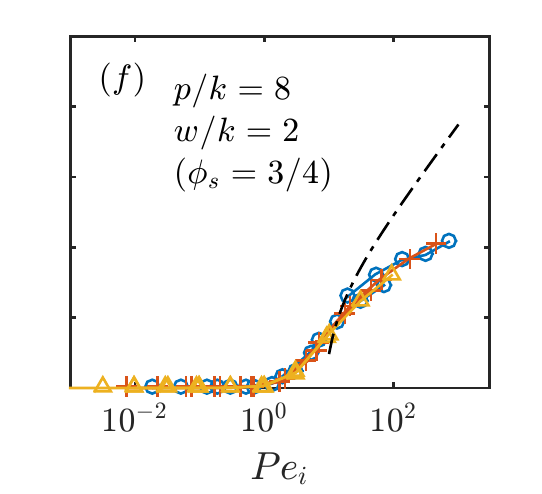}
    \captionlistentry{}
    \label{fig:NuPeOtherGeometries6}
  \end{subfigure}
  \caption{The relative contribution to the heat flux from convection for ($a$-$c$) $p/k = 4$ and ($d$-$f$) $p/k = 8$. The colours and symbols represent viscosity ratios as in fig.~\ref{fig:NuKappa}. Eq.~\eqref{eq:fit} is also shown (dashed-dotted line). The solver did not converge for some of the higher $\ReNumber_i$, so these points have been removed. }
  \label{fig:NuPeOtherGeometries}
\end{figure}

In fig.~\ref{fig:NuPeOtherGeometries}, $q_{\mathrm{conv},d}/q$ is shown for other cavity widths and pitches. For $p/k = 4$ and the same vertical wall thickness, there is a reasonable agreement to eq.~\eqref{eq:fit} (fig.~\ref{fig:NuPeOtherGeometries1}). This texture corresponds to a solid fraction $\phi_s = 1/4$ (i.e.~$1 - w/p$). For $p/k = 4$ and $\phi_s = 1/2$, it also holds (fig.~\ref{fig:NuPeOtherGeometries2}). However, for $\phi_s = 3/4$, the increase of $q_{\mathrm{conv},d}/q$ with $\PeNumber_i$ is slower (fig.~\ref{fig:NuPeOtherGeometries3}). For $p/k = 8$, the relation is valid for $\phi_s = 1/8$ and $\phi_s = 1/2$ but not for $\phi_s = 3/4$ (figs.~\ref{fig:NuPeOtherGeometries4}-f). Based on these simulations, it is clear that eq.~\eqref{eq:fit} holds at least in the interval $2 \le p/k \le 8$ for $\phi_s \le 1/2$, $10^1 < \PeNumber_i < 10^3$, $\kappa_s = \kappa_i = \kappa_\infty$, and not too thin vertical walls ($p-w\ge k$).

Assuming that $\meanDvT$ is approximately zero  above the grooves ($y\geq 0)$ -- this is confirmed in fig.~$\ref{fig:turbHeatFlux}$ and discussed more later --  we obtain, 
\begin{equation}
    \frac{q_{\mathrm{conv},d}}{q} = \frac{1}{h + 2k}\int_{-k}^h\frac{\rho c_p\meanDvT}{q}\dif y \approx \frac{1}{h + 2k}\int_{-k}^0\frac{\rho c_p\meanDvT}{q}\dif y < \frac{k}{h + 2k}.
    \label{eq:upperLimit}
\end{equation}
This inequality is based on the assumption $\rho c_p\meanDvT < q$, i.e.~$\dif \mean{T}/\dif y < 0$, which has been seen to be violated locally for high $\PrNumber_\infty$ but by a negligible amount (fig.~\ref{fig:freeFemGridRefHeatFlux}). The ratio $k/(h + 2k)$ is thus an approximate upper limit of $q_{\mathrm{conv},d}/q$, and eq.~\eqref{eq:fit} cannot be expected to be valid for $\PeNumber_i \gg 10^3$. For the laminar simulations ($h = 2k$), the limiting value is $25\%$. Moreover, from  eq.~\eqref{eq:qDecomposition}, one may see that $q_{\mathrm{conv},d}\propto k/(h + 2k)$. For the investigated groove widths ($1 \le w/k \le 7$), the vertical size of the vortices in the grooves is approximately $k$. Therefore, the integral of $\meanDvT$ scales with $k$. The definition of $q_{\mathrm{conv},d}$ then results in the scaling with $k/(h + 2k)$ (eq.~\ref{eq:qDecomposition}), which was used in eq.~\eqref{eq:fit}.



\subsection{Surface Nusselt number}
Since the dispersive convection mainly is contained in the grooves, it is possible to quantify the heat transfer through the LIS by a surface Nusselt number. We consider only the texture, equivalent to $h = 0$, and define average temperatures at the interface (the slip temperature) and the bottom of the texture, $T_s = \left.\mean{T}\right|_{y = 0}$ and $T_{lt} = \left.\mean{T}\right|_{y = -k}$, respectively. The surface Nusselt number becomes (appendix \ref{sec:surfaceNu})
\begin{equation}
    \NuNumber_i = \frac{kq}{\kappa_i (T_{lt} - T_s)} = \frac{1}{1 - q_{\mathrm{conv},d}/q}.
    \label{eq:surfaceNu}    
\end{equation}
The inequality in \eqref{eq:upperLimit}, applied to the texture only, limits $\NuNumber_i$ to finite positive values. 

Heat exchangers can be represented by thermal resistance circuits \citep{rohsenow98}. The thermal resistances of the circuit components, proportional to the inverse of their Nusselt numbers, are then evaluated separately. These resistances can then be added to form the total thermal resistance of the system if they are in series. \citet{hatte21} constructed such a model to describe convective heat transfer in pipes with LIS, with the texture and the bulk as components. However, they implicitly neglected the dispersive convection by imposing $\NuNumber_i = 1$. Since $\NuNumber_i > 1$ with convection included (eq.~\ref{eq:surfaceNu}), the resulting heat flux of the complete system is higher than their model predicts. In the upper limit of eq.~\eqref{eq:upperLimit} applied to the texture, $\NuNumber_i \rightarrow \infty$. Accordingly, the thermal resistance of the surface becomes zero. Eq.~\eqref{eq:surfaceNu}, together with eq.~\eqref{eq:fit} to describe $q_{\mathrm{conv},d}/q$, could be used for a more precise evaluation of $\NuNumber_i$ for the parameters and surface geometry considered in this study.

\section{Flow with turbulence}
\label{sec:turbulence}
\begin{figure}
  \centering
  \includegraphics[height=5.0cm]{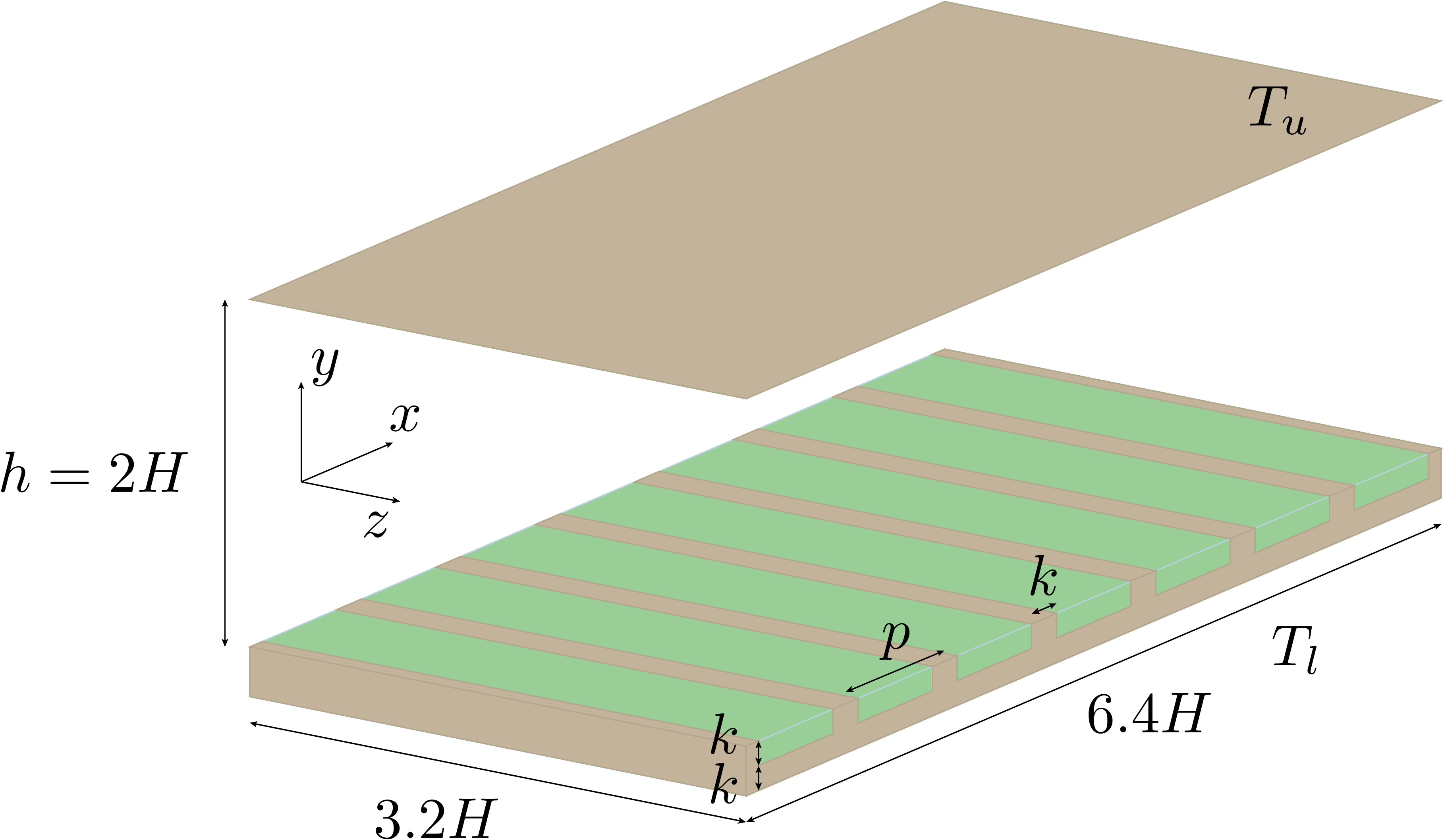}
  \caption{Sketch of the flow domain used in the turbulent simulations. The mean flow is in the positive $x$-direction. The upper wall is smooth. Transverse grooves have been added on the lower wall, on top of a slab of the same height, $k$. They correspond to a solid fraction $\phi_s = 1/4$. The colours of the infused liquid and the solid are the same as in fig.~\ref{fig:sketch}. The external fluid is not shown. }
  \label{fig:turbulentFlowConfiguration}
\end{figure}

We have carried out simulations of a turbulent channel flow with LIS using a finite difference method. The bulk Reynolds number was set to $\ReNumber_b = \rho U_b H/\mu_\infty = 2800$, resulting in a friction Reynolds number of $\ReNumber_\tau = \rho u_\tau H/\mu_\infty \approx 180$, where $H = h/2$ is the channel half-height, $U_b$ is the bulk velocity, and $u_\tau$ is the friction velocity. A constant mass flow rate was achieved by applying a uniform volume force in the infused and external liquids. The simulation domain is illustrated in fig.~\ref{fig:turbulentFlowConfiguration}. It had dimensions $(L_x, L_y, L_z) = (6.4H, 2H + 2k, 3.2H)$, and the number of grid points was $(N_x, N_y, N_z) = (640, 384, 640)$ in the streamwise, wall-normal, and spanwise directions. The grid was stretched in the wall-normal direction but uniform in the streamwise and spanwise directions. The smallest wall-normal grid spacing was $\Delta y^+ \approx 0.2$, where a superscript $+$ indicates wall units. 

We placed transverse grooves with $p/k = 4$, $\phi_s = 1/4$, and $k = 0.05H$ on the lower wall ($-0.05H \le y \le 0$), on top of a slab of solid material with the same thickness ($-0.1H \le y \le -0.05H$). The groove height corresponds to $k^{+} \approx 9$. 
All wall units were based on the friction velocity of the lower surface, computed from the balance between the applied volume force in the external flow and the wall-shear stress of the smooth upper wall. We imposed constant temperatures $T_u$ and $T_l$ at the upper and lower boundaries, respectively, similar to the laminar configuration. The viscosity ratio was set to $\mu_i/\mu_\infty = 0.4$, corresponding roughly to heptane-water (tab.~\ref{tab:properties}). The solid thermal conductivity was $\kappa_s = \kappa_i$. We performed three simulations with Prandtl numbers $\PrNumber_\infty = 1$, $2$, and $4$, respectively. These simulations were also conducted with smooth walls, replacing the grooves with solid material. For more information about the code and grid sensitivity studies, the reader is referred to \citet{ciri21}, where the same code was used with a similar setup. The results from the turbulent simulations are summarised in tab.~\ref{tab:turbResults} and will be discussed in the following subsections. Some statistics from the turbulent simulations are given in appendix \ref{sec:numericalMethodsTurbulence}.

\begin{table}
\centering
\begin{tabular}{lllllllllll}
$\PrNumber_\infty$ & $\DR$ (\%) & $\dfrac{q-q^0}{q^0}$ (\%)& $Nu^0$ & $\epsilon\cdot10^{3}$ & $\dfrac{q_{\mathrm{conv},d}}{q}\cdot10^{3}$ & $\dfrac{q}{q^0}\dfrac{\tau_0}{\tau}$ \vspace{0.2cm} \\ 
1 & 2.8 & 3.3  & 5.8 & 1.7 & 3.8 & 1.06 \\
2 & 2.8 & 7.8  & 7.5 & 2.3 & 7.3 & 1.11\\
4 & 2.8 & 13.6 & 9.3 & 2.4 & 10.5 & 1.17 \\
\end{tabular}
\caption{Summary of results from the turbulent simulations. The friction Reynolds number of the smooth channel flow was $\ReNumber_\tau^0 = 178.7$, and for the flow with LIS, $\ReNumber_\tau^0 = 176.1$, based on the friction velocity of the textured surface. The drag reduction ($\DR$) was computed by eq.~\eqref{eq:turbulentDR} and the quantities $\tau^0$, $q^0$, and $Nu^0$ are wall-shear stress, heat flux and Nusselt number of the smooth-wall simulations, respectively. The heat transfer increase has an uncertainty of $\pm 2\%$ (appendix \ref{sec:numericalMethodsTurbulence}).}
\label{tab:turbResults}
\end{table}

\subsection{Turbulent heat transfer mechanisms}

\begin{figure}
  \centering
  \includegraphics[width=11cm]{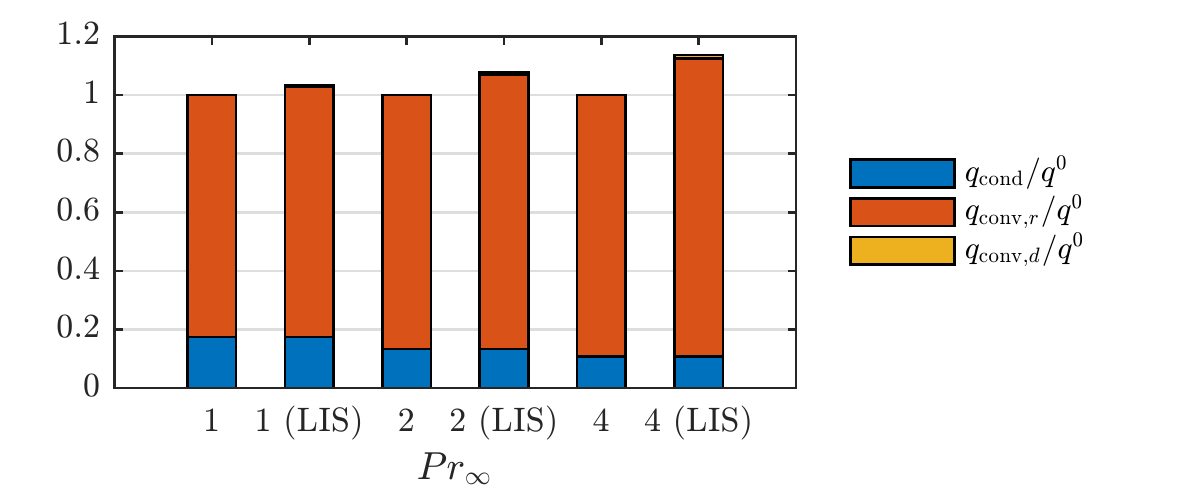}
  \caption{Contributions to $q/q^0$ for the different turbulent flow cases. On this scale, $q_{\mathrm{conv},d}$ is hardly visible, corresponding to around $1\%$ or less of $q$ (tab.~\ref{tab:turbResults}). }
  \label{fig:qBudgetTurb}
\end{figure}

All the terms in eq.~\eqref{eq:qDecomposition} contribute to the heat flux in a turbulent flow since the flow contains random fluctuations. The change in the heat flux can be written as, 
\begin{equation}
    \frac{q}{q^0} = \frac{q_{\mathrm{cond}}}{q^0} + \frac{q_{\mathrm{conv},r}}{q^0} + \frac{q_{\mathrm{conv},d}}{q^0}.
    \label{eq:HTETurbulenceDefinition}
\end{equation}
%
%
Figure \ref{fig:qBudgetTurb} shows the heat flux components for smooth and liquid-infused surfaces. 
We observe that the contribution of the dispersive term is very small ($q_{\textrm{conv,d}}/q^0\sim 1\%$) for all Prandtl numbers. Nevertheless, the total increase of the heat flux compared to the smooth-wall flow, $q/q^0$, was $3.3\%$, $7.8\%$, and $14\%$ for $\PrNumber_\infty = 1$, $2$, and $4$, respectively (tab.~\ref{tab:turbResults}). As shown in fig.~\ref{fig:qBudgetTurb}, for all these cases, the increase in $q_{\mathrm{conv},r}/q^0$ dominates the change in the heat flux. However, the main reason behind the enhancement of $q_{\mathrm{conv},r}/q^0$ is the small but finite value of $q_{\mathrm{conv},d}/q$, as will be shown later in this section.

\begin{figure}
  \centering
  \begin{subfigure}{0.45\textwidth}
     \includegraphics[width=5.5cm]{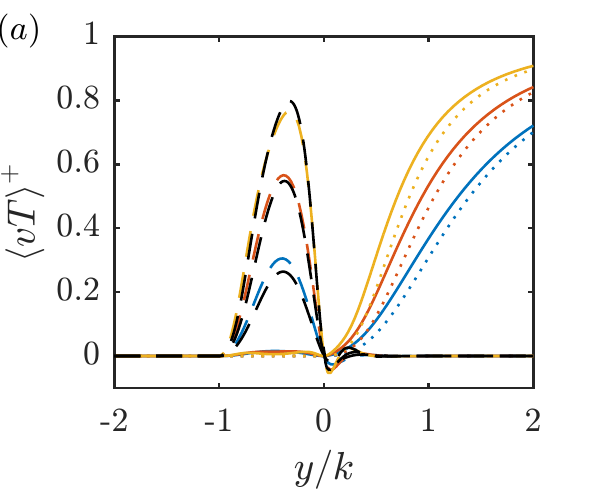}
    \captionlistentry{}
    \label{fig:turbHeatFlux}
  \end{subfigure}
  \begin{subfigure}{0.45\textwidth}
    \includegraphics[width=5.5cm]{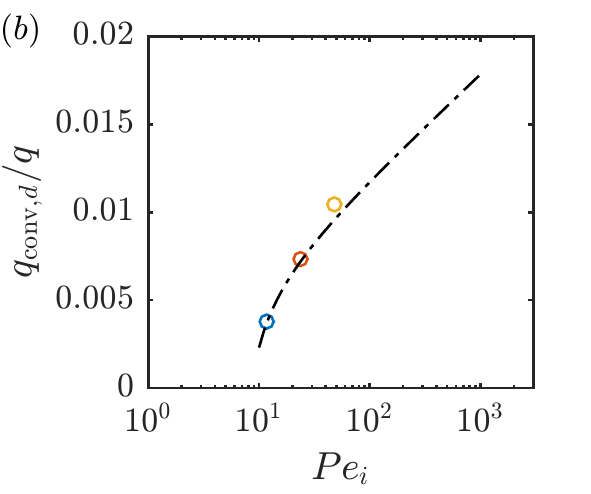}
    \captionlistentry{}
    \label{fig:turbGrooveVortexHeat}
  \end{subfigure}
  \caption{($a$) Comparison of convective heat flux for LIS in laminar flow (\longbroken, black) and turbulent flow (\longbroken, $\meanDvT^+$; \full, $\left<v'T'\right>^+$; \dotted, smooth-wall reference; blue, red, and yellow for $\PrNumber_\infty = 1$, $2$, and $4$, respectively). For laminar and turbulent flows, $\meanDvT^+ \approx 0$ in the bulk flow ($y > 0$), and for the turbulent flow, $\left<v'T'\right>^+ \approx 0$ inside the grooves ($y < 0$). ($b$) Comparison of $q_{\mathrm{conv},d}/q$ from the turbulent simulations (colours as in $a$) and eq.~\eqref{eq:fit} (dash-dotted line). }
\end{figure}

First, however, we note that if eq.~(\ref{eq:upperLimit}) is valid also for turbulent flows, then the upper limit of $q_{\mathrm{conv},d}/q$ for the turbulent flow setup is $2.4\%$. For a corresponding symmetric channel, the upper limit would be $4.5\%$. The potential contribution from $q_{\mathrm{conv},d}/q$ is, therefore, somewhat restricted for this setup. 
Fig.~\ref{fig:turbHeatFlux} shows $\rho c_p\meanDvT/q = \meanDvT^+$ from turbulent and laminar simulations for the same Reynolds number $\ReNumber_i = 29$. The agreement between the two systems is 
reasonably good and we can thus use the laminar flow to interpret the turbulent simulations.  This is also corroborated by fig.~\ref{fig:turbGrooveVortexHeat}, where $q_{\mathrm{conv},d}/q$ is compared to eq.~\eqref{eq:fit}, corresponding to fig.~\ref{fig:NuPeOtherGeometries1} for the laminar simulations. 
Indeed, earlier studies indicate that dispersive quantities of rough-wall flow can be reproduced by laminar flow if the offset of the mean velocity in the logarithmic region compared to smooth-wall flow is small ($\Delta U^+ \lesssim 2$) \citep{abderrahaman-elena19}. Correspondence between the laminar and turbulent results of $\meanDvT^+$ is therefore expected here. 

%

%

In the following, we rewrite eq.~\eqref{eq:HTETurbulenceDefinition} to gain a better insight into the role of the random fluctuations in the heat transfer process. The conduction term corresponds to
\begin{equation}
     \frac{q_\mathrm{cond}}{q^0} = \frac{q_\mathrm{cond}^0}{q^0} = \frac{\kappa_\infty (T_l - T_u)}{(h+2k)q^0} = \frac{1}{\NuNumber^0},
     \label{eq:qcondNu}
\end{equation}
where $\NuNumber^0$ is the Nusselt number of the smooth-wall flow.
Fig.~\ref{fig:turbHeatFlux} compares the convective heat flux of a smooth and liquid-infused surface, where we observe only minor changes in random component, $\mean{v'T'}^+$. We introduce the difference in $q_{\mathrm{conv},r}/q$ between the flow over the LIS and in the smooth channel, 
\begin{equation}
    \epsilon = \frac{q_{\mathrm{conv},r}}{q} - \frac{q_{\mathrm{conv},r}^0}{q^0}.
    \label{eq:epsilon}
\end{equation}
It follows that
\begin{equation}
    \frac{q_{\mathrm{conv},r}}{q^0} = \frac{q_{\mathrm{conv},r}^0}{q^0}\frac{q}{q^0} + \epsilon\frac{q}{q^0} = \left(1 - \frac{1}{\NuNumber^0} \right)\frac{q}{q^0} + \epsilon\frac{q}{q^0},
    \label{eq:randomConvectionq0}
\end{equation}
using $q_{\mathrm{conv},r}^0/q^0 = 1 - q_{\mathrm{cond}}^0/q^0$ and eq.~\eqref{eq:qcondNu} in the second step. Assuming $\epsilon = 0$, the ratio of the random convection to the total heat flux, $q_{\mathrm{conv},r}/q$, does not change for the flow over LIS compared to the smooth-wall flow. Accordingly, $q_{\mathrm{conv},r}/q^0$ is proportional to $q/q^0$. 

Now, using eqs.~\eqref{eq:qcondNu} and \eqref{eq:randomConvectionq0}, eq.~\eqref{eq:HTETurbulenceDefinition} can be written as,
\begin{equation}
    \frac{q}{q^0} = 1 + \NuNumber^0\left(\frac{q_{\mathrm{conv},d}}{q^0} + \epsilon\frac{q}{q^0}\right).
    \label{eq:HTETurbulenceDefinition3}
\end{equation}
Assuming $\epsilon = 0$, we have virtually eliminated $q_{\mathrm{conv},r}/q^0$ and $q_\mathrm{cond}/q^0$ from eq.~\eqref{eq:HTETurbulenceDefinition}; instead $\NuNumber^0$ acts as amplification factor for $q_{\mathrm{conv},d}/q^0$. For a laminar flow $\NuNumber^0 = 1$, i.e. there is no amplification.
On the other hand, if there are random fluctuations in the flow, $\NuNumber^0 = q_{\mathrm{conv},r}^0/q_\mathrm{cond}^0 + 1$, demonstrating how $\NuNumber^0$ increases due to convection. Hence, $\NuNumber^0$ is a measure of the amplification of $q_{\mathrm{conv},d}/q^0$ by the random fluctuations. Indeed, this amplification can be substantial. Values measured from the simulations with smooth walls were $\NuNumber^0 = 5.8$, $7.5$, and $9.3$ for $\PrNumber_\infty = 1, 2$, and $4$, respectively (tab.~\ref{tab:turbResults}). This effect can be expected to increase with increasing bulk Reynolds or Prandtl numbers; \citet{kim20} noticed that $\NuNumber^0$ increases linearly with $\ReNumber_\tau$ for a channel with smooth isothermal walls. As $q/q^0$ increases due to the dispersive convection, so does $q_{\mathrm{conv},r}/q^0$ by eq.~\eqref{eq:randomConvectionq0}.

\subsection{Predictions of heat transfer increase}
By rearranging eq.~\eqref{eq:HTETurbulenceDefinition3}, we get an analytical expression for the change in heat flux as
\begin{multline}
    \frac{q}{q^0} = \left [{1 - \NuNumber^0\left(\frac{q_{\mathrm{conv},d}}{q} + \epsilon\right)}\right ]^{-1} = \left [1 - \NuNumber^0\left (\frac{q_{\mathrm{conv},d}}{q}\right) \right]^{-1} + O(\epsilon)\\ = \left [{1 - \frac{q^0k}{\kappa_\infty (T_l - T_u)}0.11\ln(\PeNumber_i - 7.6)}\right ]^{-1} + O(\epsilon).
    \label{eq:heatFluxTurb}
\end{multline}
Here, we have performed a Taylor series expansion around $\epsilon = 0$. Eq.~\eqref{eq:fit} has been used in the latter form of the expression (applicable if $10^1 < \PeNumber_i < 10^3$). 
The Nusselt number $\NuNumber^0$ (and $q^0$) depends on $\PrNumber_\infty$ and $\ReNumber_b$ but can be determined without performing simulations of LIS since it is a result of the smooth-wall flow. The other input parameter, $\ReNumber_i$, can be computed or estimated as in fig.~\ref{fig:re_relation_slip_lengths}. Therefore, if $\epsilon$ is neglected, eq.~\eqref{eq:heatFluxTurb} predicts the heat flux of the LIS. \citet{rastegari15} and \citet{ciri21} constructed similar relationships for drag reduction and heat transfer increase with isothermal solids, respectively. 

\begin{figure}
  \centering
  \includegraphics[width=5.5cm]{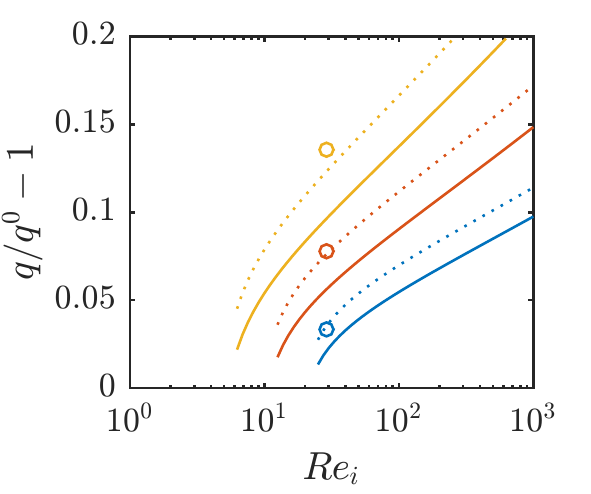}
  \caption{Comparison of eq.~\eqref{eq:heatFluxTurb}, assuming $\epsilon = 0$ (solid lines) and $\epsilon = 2.3\cdot 10^{-3}$ (dotted lines, corresponding to $\PrNumber_\infty = 2$), to simulation results shown with circles for the three values of $\PrNumber_\infty$. The colours are the same as in fig.~\ref{fig:turbHeatFlux}. }
  \label{fig:heatFluxTurb}
\end{figure}

Eq.~\eqref{eq:heatFluxTurb} is shown in fig.~\ref{fig:heatFluxTurb}, both for $\epsilon = 0$ and $\epsilon = 2.3\cdot10^{-3}$. The latter corresponds to the measured value for $\PrNumber_\infty = 2$. For $\PrNumber_\infty = 1$, $\epsilon$ was somewhat lower, and for $\PrNumber_\infty = 4$ slightly higher (see tab.~\ref{tab:turbResults}). The positive values of $\epsilon$ further enhance the heat flux, which is seen from the first form of the expression in eq.~\eqref{eq:heatFluxTurb}. This expression with $\epsilon = 0$ is thus a lower limit of $q/q^0$. Since the sum of $\epsilon$ and $q_{\mathrm{conv},d}/q$ enters the equation, their magnitudes can be compared: $q_{\mathrm{conv},d}/q$ is $2.2$, $3.1$, and $4.4$ times larger for $\PrNumber_\infty = 1$, $2$, and $4$, respectively.

\begin{figure}
  \centering
  \begin{subfigure}{0.45\textwidth}
     \includegraphics[width=5.5cm]{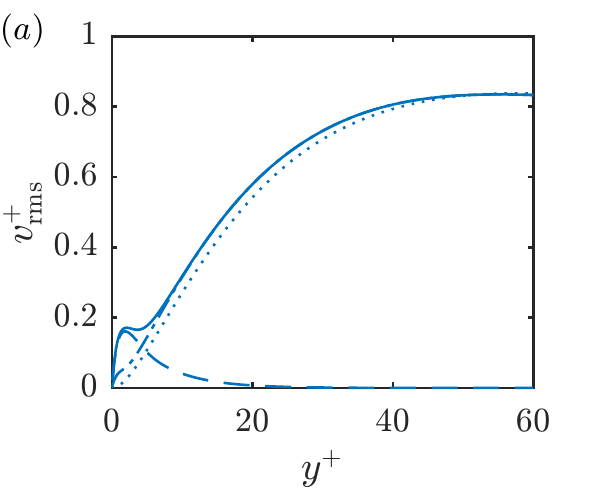}
    \captionlistentry{}
    \label{fig:turbVRmsSurface}
  \end{subfigure}
  \begin{subfigure}{0.45\textwidth}
    \includegraphics[width=5.5cm]{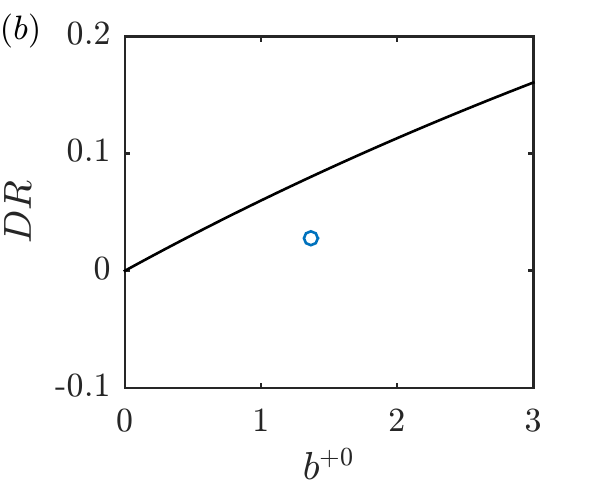}
    \captionlistentry{}
    \label{fig:rastegariAkhavanDR}
  \end{subfigure}
  \caption{($a$) The root-mean-squared wall-normal velocity for turbulent flow over LIS (\full, $v_\mathrm{rms}^+$; \longbroken, dispersive component $v_{\mathrm{rms},d}^+$; \chain, random component $v_{\mathrm{rms},r}^+$; \dotted, smooth-wall reference). ($b$) The drag reduction of the current LIS as a function of the slip length in nominal wall units (circle), compared to the ideal relation by \citet{rastegari15} (solid line). }
  \label{fig:productionQuantities}
\end{figure}

\subsection{Heat flux to drag ratio}
The slight increase of $\mean{v'T'}^+$ is reflected by the wall-normal velocity fluctuations. As shown in fig.~\ref{fig:turbVRmsSurface}, the root-mean-squared value $v_\mathrm{rms}^+$ increases near the surface. The varying slip/no-slip condition at $y = 0$, acting like roughness, causes this effect. 
Similar roughness effects severely reduce the drag benefits for transverse grooves \citep{ciri21}. The LIS gave a slight drag reduction compared to the smooth wall,
\begin{equation}
    \DR = \frac{\tau^0 - \tau}{\tau^0} = 0.028,
    \label{eq:turbulentDR}
\end{equation}
where $\tau$ is the total drag of the lower surface, and $\tau^0$ is the wall-shear stress from the reference simulation with smooth walls.
The current $\DR$ and the ideal drag reduction relationship by \citet{rastegari15} are shown in fig. \ref{fig:rastegariAkhavanDR}. This relationship was derived for channels with symmetric walls but can also be applied to asymmetric configurations if $\DR$ is computed by eq.~\eqref{eq:turbulentDR} \citep{fu17, ciri21}. Without roughness effects, it predicts that the drag reduction would be 8\%. Nevertheless, we achieve $\DR > 0$ together with a heat transfer increase for this geometry. Interfaces deformations, which has been neglected here, tend to reduce $\DR$ further \citep{cartagena18, ciri21, sundin21}.

The heat transfer efficiency of the system can be measured by the heat flux to drag ratio or, equally, the Reynolds analogy factor, $2\StNumber/C_f$, where $\StNumber$ is the Stanton number and $C_f$ is the friction coefficient. According to the Reynolds analogy, $2\StNumber/C_f = 1$ for flow over smooth walls with $\PrNumber_\infty = 1$ \citep{kestin63}. Changes in the heat flux to drag ratio when introducing surface modifications at this Prandtl number thus indicate a breakage of the Reynolds analogy. A growth or a reduction equal increased or decreased heat transfer efficiency, respectively. However, the exact value of $2\StNumber/C_f$ for smooth-wall flows depends on the normalisation used to form the non-dimensional numbers. 
The heat flux to drag ratio normalised with the smooth-wall reference values, $(q/q^0)(\tau_0/\tau)$, is a valid measure of the heat transfer efficiency independently of the Prandtl number. This quantity is reported in tab.~\ref{tab:turbResults}. Since both $q/q^0 > 1$ and $\tau_0/\tau > 1$ for the current setup, $(q/q^0)(\tau_0/\tau)$ exceeds unity, having a maximum value of $1.17$ for $\PrNumber_\infty = 4$. For LIS and SHS with isothermal solids, it ranges between $0.9$ and $1.2$ \citep{ciri21}. Rough walls with either irregular \citep{forooghi18} or structured textures (such as transverse bars or rods \citep{leonardi15}) have been seen to reduce the Reynolds analogy factor.

\section{Conclusions}
\label{sec:conclusions}
For LIS with transverse grooves, shear stress due to the external flow induces recirculation in the cavities. This recirculation increases the heat transfer by dispersive convection, contributing $q_{\mathrm{conv},d}$ to the total heat flux. A net increase in the heat flux compared to a smooth surface, $q/q^0 > 1$, can be achieved in laminar flows if the thermal conductivity of the solid is similar to or smaller than the conductivity of infused liquid, $\kappa_s \lesssim \kappa_i$. If $\kappa_s \approx \kappa_i$, $q_{\mathrm{conv},d}$ may be essential. Therefore,  we have investigated the convection for $\kappa_s =\kappa_i$ in greater detail. The ratio $q_{\mathrm{conv},d}/q$ can be expressed as a function of the Péclet number based on the slip velocity and the groove height, $\PeNumber_i$. This function resembles a logarithmic function in the interval $10^1 < \PeNumber_i < 10^3$, eq.~\eqref{eq:fit}. The same approximation is also applicable for turbulent external flow. 

Dispersive convection occurs only inside and in the vicinity of the surface. Therefore, apart from the dependency on $\PeNumber_i$, $q_{\mathrm{conv},d}$ is proportional to the ratio of the groove height to the total system height. This ratio is also an approximate upper limit of $q_{\mathrm{conv},d}/q$ (eq.~\ref{eq:upperLimit}). For turbulent flows, the convection from random fluctuations in the bulk flow, $q_{\mathrm{conv},r}$, dominates. However, $q_{\mathrm{conv},r}/q$ changes only slightly from the smooth-wall flow. Therefore, the dispersive convection can amplify the heat flux considerably even for turbulent flows, as expressed by eq.~\eqref{eq:heatFluxTurb}. 

The convection in the LIS texture could be used in applications to achieve increased heat transfer. Numerous liquids can be used for LIS, whereas constraints on solid surface materials, such as weight, could limit their thermal conductivity. Low weight, for example, has been a driving force in the increased usage of micro heat exchangers \citep{dixit15}. In addition to the increased surface heat flux, the slip induced by the LIS leads to reduced friction which lowers the required power input \citep{hatte21}. These two properties make LIS augmented convective heat transfer a promising method for achieving more efficient heat exchangers.

\backsection[Acknowledgements]{
We are thankful to U\v{g}is L\={a}cis for his help with the setup in FreeFem++. The support of this work from SSF, the Swedish Foundation for Strategic Research (Future Leaders grant FFL15:0001), is greatly acknowledged. M.~H.~acknowledges the generous support from The Wenner-Gren Foundations.}

\backsection[Declaration of interests]{The authors report no conflict of interest.}


\appendix

\section{FIK identities}
\label{sec:fikIdentities}
We begin with the energy equation, eq.~\eqref{eq:dim_heat_equation}, and assume periodicity in the streamwise and spanwise directions. Taking the average in time and the streamwise and spanwise directions, denoted by the operator $\mean{}$,
\begin{equation}
    \rho c_p\frac{\dif }{\dif y}\mean{vT} - \frac{\dif }{\dif y}\mean{\kappa\frac{\partial T}{\partial y}} = 0.
    \label{eq:averagedHeatEq}
\end{equation}
Integration from $-2k$, corresponding to the bottom of the domain, to $y$, expresses a balance between convective fluxes and conduction,
\begin{equation}
  \rho c_p\mean{vT} - \rho c_p\left.\mean{vT}\right|_{-2k} - \mean{\kappa\frac{\partial T}{\partial y}} + \left.\mean{\kappa\frac{\partial T}{\partial y}}\right|_{-2k} = 0.
\end{equation}
We integrate a second time from $-2k$ to $h$,
\begin{multline}
    \rho c_p\int_{-2k}^h\mean{vT}\dif y - \rho c_p\left.\mean{vT}\right|_{-2k}(h + 2k) \\ -\int_{-2k}^h\mean{\kappa \frac{\partial T}{\partial y}}\dif y + \kappa\left.\mean{\frac{\partial T}{\partial y}}\right|_{-2k}(h + 2k) = 0.
    \label{eq:secondFIKIntegral}
\end{multline}
Identifying the heat flux at the bottom of the domain, 
\begin{multline}
    q = -\kappa\left.\mean{\frac{\partial T}{\partial y}}\right|_{-2k} + \rho c_p\left.\mean{vT}\right|_{-2k} =\\ -\frac{1}{h + 2k}\int_{-2k}^h\mean{\kappa \frac{\partial T}{\partial y}}\dif y + \frac{\rho c_p}{h + 2k}\int_{-2k}^h\mean{vT}\dif y.
\end{multline}
The integral of the conduction term can be split into an external fluid and a surface term by considering integration limits from $y = 0$ to $h$ and $y = -2k$ to $0$, respectively. We decompose the convective flux (see appendix \ref{sec:flowDecomp}),
\begin{equation}
    \mean{vT} = \meanDvT + \mean{v'T'},
\end{equation}
where $\meanDvT$ is the dispersive component and $\mean{v'T'}$ is the random, external flow component. It results in
\begin{multline}
    q = - \frac{1}{h + 2k}\int_{0}^h\mean{\kappa \frac{\partial T}{\partial y}}\dif y - \frac{1}{h + 2k}\int_{-2k}^0\mean{\kappa \frac{\partial T}{\partial y}}\dif y + \frac{\rho c_p}{h + 2k}\int_{-k}^h\mean{v'T'}\dif y \\+ \frac{\rho c_p}{h + 2k}\int_{-k}^h\meanDvT\dif y = 
    q_{\mathrm{cond},\infty} + q_{\mathrm{cond},s} + q_{\mathrm{cond},i} + q_{\mathrm{conv},r} + q_{\mathrm{conv},d}.
    \label{eq:qDivision}
\end{multline}
where conductive and convective parts have been identified. The conduction in the solid and the infused liquid are separated after the equal sign. They can be written as 
\begin{align}
    q_{\mathrm{cond},s} &= -\frac{1}{h + 2k}\int_{-2k}^0\mean{\kappa_s \frac{\partial T}{\partial y}\chi_s}\dif y, \\ q_{\mathrm{cond},i} &= -\frac{1}{h + 2k}\int_{-2k}^0\mean{\kappa_i \frac{\partial T}{\partial y}(1 - \chi_s)}\dif y,
\end{align}
where the indicator function $\chi_s = 1$ inside the solid and $0$ in the liquid. The thermal conductivity $\kappa$ has been identified with its value in each domain. It might therefore be moved out of the average operators and the integrals. Eq.~\eqref{eq:qDivision} then results in eq.~\eqref{eq:qDecomposition}. 

\subsection{Laminar flow over smooth wall}
\label{sec:fikIdentitiesLaminar}
For a laminar flow over a smooth wall, eq.~\eqref{eq:averagedHeatEq} reduces to 
\begin{equation}
    -\frac{\dif }{\dif y}\mean{\kappa\frac{\partial T}{\partial y}} = 0,
    \label{eq:averagedHeatEqSmoothLaminar} 
\end{equation}
which means $-\mean{\kappa \partial T/\partial y}$ is constant everywhere, equal to $q$. Integrating eq.~\eqref{eq:averagedHeatEqSmoothLaminar} over the solid and the external fluid,
\begin{equation}
    q = \frac{\kappa_s}{2k}(T_l - T_s) \quad \text{and} \quad q = \frac{\kappa_\infty}{h}(T_s - T_u),
    \label{eq:smoothLaminarInterfTemp}
\end{equation}
respectively, where $T_l = \left.\mean{T}\right|_{y=-2k}$, $T_u = \left.\mean{T}\right|_{y=h}$, and $T_s = \left.\mean{T}\right|_{y=0}$. These two expressions give a solution for $T_s$,
\begin{equation}
    T_s = \left(\dfrac{\kappa_\infty}{h}T_u + \dfrac{\kappa_s}{2k}T_l\right) \bigg/ \left(\dfrac{\kappa_\infty}{h} + \dfrac{\kappa_s}{2k}\right).
\end{equation}
Putting this expression back into eq.~\eqref{eq:smoothLaminarInterfTemp},
\begin{equation}
    q = \frac{T_l - T_u}{2k/\kappa_s + h/\kappa_\infty}.
    \label{eq:smoothWallq}
\end{equation}

\subsection{Homogeneous thermal conductivity}
\label{sec:fikIdentitiesHomogeneousConductivity}
Simplifications of eq.~\eqref{eq:qDivision} can be made by assuming $\kappa = \kappa_i = \kappa_\infty = \kappa_s$. Then,
\begin{multline}
    - \frac{1}{h + 2k}\int_{0}^h\mean{\kappa \frac{\partial T}{\partial y}}\dif y - \frac{1}{h + 2k}\int_{-2k}^0\mean{\kappa \frac{\partial T}{\partial y}}\dif y \\ = \frac{\kappa}{h + 2k}(T_s - T_u) + \frac{\kappa}{h + 2k}(T_l - T_s) = \frac{\kappa}{h + 2k}(T_l - T_u).
\end{multline}
Hence,
\begin{equation}
    q = \frac{\kappa}{h + 2k}(T_l - T_u) + \frac{\rho c_p}{h + 2k}\int_{-k}^h\mean{v'T'}\dif y + \frac{\rho c_p}{h + 2k}\int_{-k}^h\meanDvT\dif y.
\end{equation}
The first term is equal to eq.~\eqref{eq:smoothWallq}, as it corresponds to only the conduction.

\section{Flow decomposition}
\label{sec:flowDecomp}
For the quantity $u$,
\begin{equation}
    U = \mean{u} = \frac{1}{L_x L_z \Delta t}\int_0^{L_x}\int_0^{L_z}\int_{t'}^{t'+\Delta t} u(x,y,z,t) \dif t \dif z \dif x,
\end{equation}
starting averaging at some time $t'$ and over a time difference $\Delta t$. The extent of the domain in the streamwise and spanwise directions is represented by $L_x$ and $L_z$, respectively. We may also introduce an average in time with a zero plane average, $\meanR{ }~$, such that
\begin{equation}
    \meanR{u} = \frac{1}{\Delta t}\int_{t'}^{t'+\Delta t} u(x,y,z,t) \dif t - \mean{u},
\end{equation}
giving the dispersive fluctuations. Since the transverse grooves are aligned in the spanwise direction, spatial averaging in this direction may be applied when computing these components. The periodicity of the structures can also be exploited. Using a triple decomposition (neglecting non-linear terms, see \citet{abderrahaman-elena19}), 
\begin{equation}
    u = U + \meanR{u} + u',
\end{equation}
where $u'$ are the random fluctuations attributed to the background turbulence. The root-mean-squared value is $u_\mathrm{rms} = \sqrt{\mean{(u-U)^2}}$. The dispersive root-mean-squared value is $u_{\mathrm{rms},d} = \sqrt{\mean{\meanR{u}^2}}$. Also,
\begin{equation}
    u_\mathrm{rms}^2 = \mean{(\meanR{u} + u')^2} = u_{\mathrm{rms},d}^2 + u_{\mathrm{rms},r}^2 + 2\mean{\meanR{u}u'},
\end{equation}
where $u_{\mathrm{rms},r} = \sqrt{\mean{u'^2}}$. The last term is zero since there is no correlation between $\meanR{u}$ and $u'$. Such cross-correlations are always zero because $\meanR{u}$ is independent of $t$, whereas the average of $u'$ in $t$ is zero. Similarly, for wall-normal velocity and temperature,
\begin{equation}
    \mean{vT} = \mean{(\meanR{v} + v')(\mean{T} + \meanR{T} + T')} = \meanDvT + \mean{v'T'}.
\end{equation}

\section{Simulations with FreeFem++}
\label{sec:numericalMethodsFreeFem}
The finite element solver FreeFem++ \citep{hect12} was used to carry out the laminar simulations of the unit cell in two dimensions. We used a built-in tool to generate a triangle mesh with similar spacing in the streamwise and wall-normal direction, conforming to the solid boundaries and the interface. The velocities and the temperature were described by quadratic ($P_2$) and the pressure by linear ($P_1$) finite elements. The momentum and continuity equations were solved simultaneously in an iterative manner until convergence was reached. Afterwards, the equation for the temperature was solved.

\begin{figure}
    \centering
    \begin{subfigure}{0.45\textwidth}
        \includegraphics[width=5.5cm]{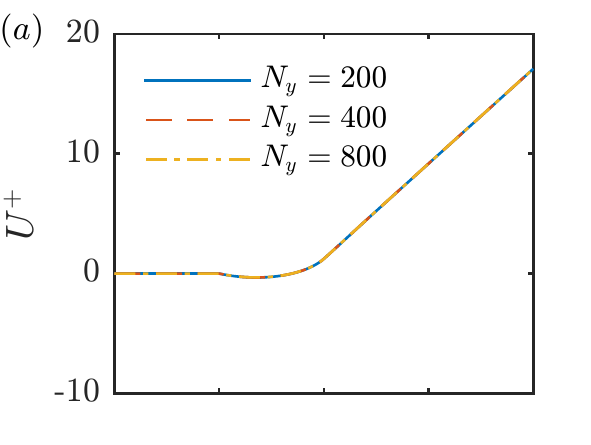}
        \captionlistentry{}
    \end{subfigure}
    \begin{subfigure}{0.45\textwidth}
        \includegraphics[width=5.5cm]{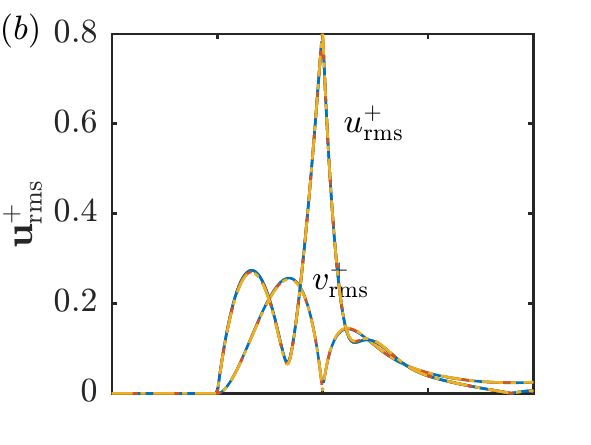}
        \captionlistentry{}
    \end{subfigure}
    \begin{subfigure}{0.45\textwidth}
        \includegraphics[width=5.5cm]{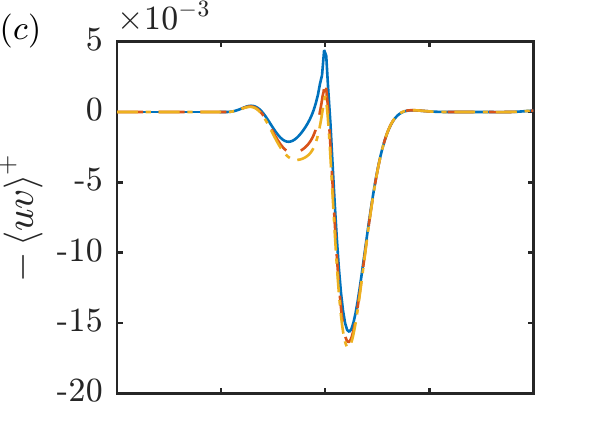}
        \captionlistentry{}
    \end{subfigure}
    \begin{subfigure}{0.45\textwidth}
        \includegraphics[width=5.5cm]{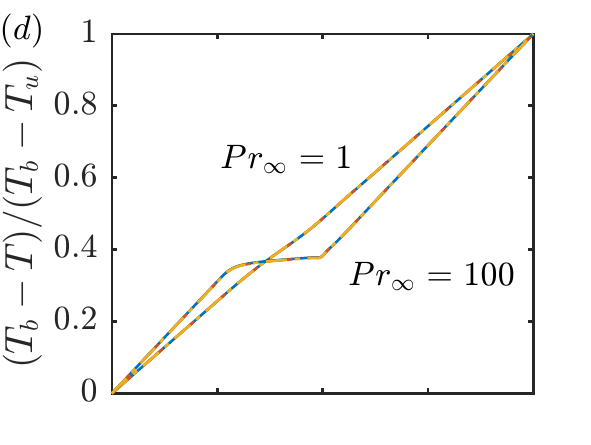}
        \captionlistentry{}
        \label{fig:freeFemGridRefMeanTemp}
    \end{subfigure}
    \begin{subfigure}{0.45\textwidth}
        \includegraphics[width=5.5cm]{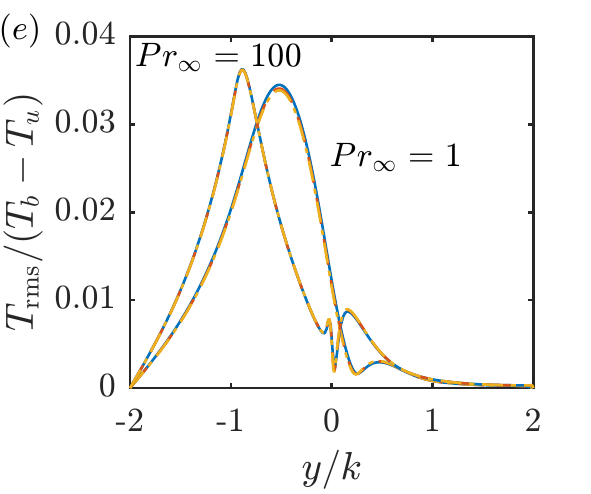}
        \captionlistentry{}
    \end{subfigure}
    \begin{subfigure}{0.45\textwidth}
        \includegraphics[width=5.5cm]{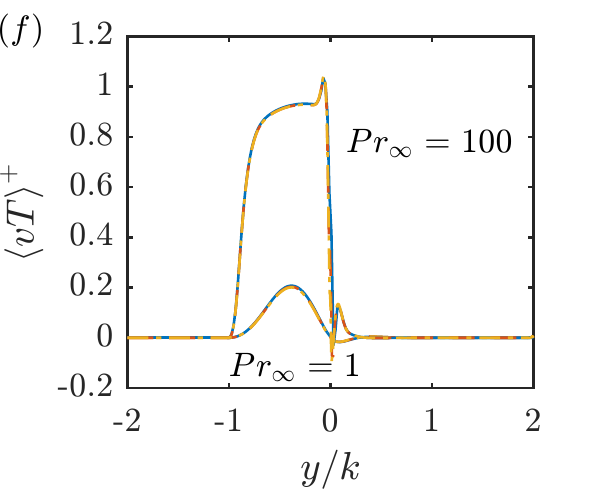}
        \captionlistentry{}
        \label{fig:freeFemGridRefHeatFlux}
    \end{subfigure}
    \caption{Grid convergence study for laminar flow with FreeFem++, with $\kappa_i = \kappa_\infty = \kappa_s$, $p/k = 4$, $\mu_i/\mu_\infty = 0.4$, $\ReNumber_i = 24$, and $\PrNumber_\infty = 1$ and $100$. Three grids were used, with $N_y = 200$, $400$, and $800$ cells in the wall-normal direction. The horizontal scale is normalised with groove height, $k$, and corresponds to the complete domain, with the interface at $y = 0$. In ($b$), $u_\mathrm{rms}^+$ and $v_\mathrm{rms}^+$ denote the streamwise and wall-normal root-mean-squared velocities, respectively.}
    \label{fig:freeFemGridRefinement}
\end{figure}

We performed a grid convergence study for $\kappa_i = \kappa_\infty = \kappa_s$, $p/k = 4$, $w/k = 3$ ($\phi_s = 1/4$), and $\mu_i/\mu_\infty = 0.4$, corresponding to the values used in the turbulent simulations. Further, $\ReNumber_\infty = 270$, $\ReNumber_i = 24$, and both $\PrNumber_\infty = 1$ and $100$ were tested. The simulations were run with grids generated with $N_y = 200$, $400$, and $800$ triangles along the wall-normal boundary. The differences in $U_\infty$ and $q$ between the grids were less than $0.1 \%$ for $\PrNumber_\infty = 1$ and $100$. These insignificant differences show that the simulations, to a large extent, are independent of the grid. Plots of relevant quantities are shown in fig.~\ref{fig:freeFemGridRefinement}. Velocities are normalised with the friction velocity, $u_\tau$, temperature with $T_l - T_u$, and $\mean{vT}$ with $q/(\rho c_p)$. The heat flux, $q$, was assessed at the top boundary. For $\PrNumber_\infty = 100$, there was a noticeable peak in the total heat flux at the interface, decreasing with increased resolution. It indicates the need for a fine grid to resolve the gradient in $\mean{vT}$ (fig.~\ref{fig:freeFemGridRefHeatFlux}). However, because of the consistency in the heat flux elsewhere, all three resolutions were still considered acceptable. For the rest of the simulations, we used $N_y = 400$.

\section{Analytical slip length in the Stokes limit}
\label{sec:analyticalSlipLength}
For constant shear stress, the velocity at $y = h$ is $U_\infty = h \left.\dif U/\dif y\right|_{y = 0} + U_s = (h + b)\left.\dif U/\dif y\right|_{y = 0}$. Hence,
\begin{equation}
    \frac{\ReNumber_i}{\ReNumber_\infty} = \frac{b}{h + b}\frac{k\mu_\infty}{h\mu_i}.
\end{equation}

An expression for the slip length in the Stokes limit of both longitudinal and transverse LIS was derived by \citet{schonecker14}. For transverse grooves, the slip length normalised by the pitch is
\begin{equation}
    \frac{b}{p} = \frac{\ln\left(\cos\left(\dfrac{\pi a}{2}\right)\right)}{2\pi + \dfrac{\mu_i}{2aD_t(A,a)\mu_\infty}\ln\left(\dfrac{1 + \sin\left(\dfrac{\pi a}{2}\right)}{1 - \sin\left(\dfrac{\pi a}{2}\right)}\right)},
\end{equation}
where $a = w/p$ is the liquid fraction (i.e.~$1-\phi_s$), and $D_t(A,a)$ is the maximum value of the local slip length normalised by $w$ and $\mu_\infty/\mu_i$, with $A = k/w$ here being the groove aspect ratio. It can be expressed as
\begin{equation}
    D_t(A,a) = f(a)\beta~\mathrm{erf}\left(\dfrac{g(a)\sqrt{\pi}}{8f(a)\beta}A\right),
\end{equation}
where $\beta = 0.505/(2\pi)$, $\mathrm{erf}(x)$ is the error function,
\begin{equation}
    f(a) = -\frac{\ln\left(\dfrac{1 + \sin\left(\dfrac{\pi a}{2}\right)}{1 - \sin\left(\dfrac{\pi a}{2}\right)}\right)}{2a\ln 2 \left(1 + \dfrac{2\ln\left(\cos\left(\dfrac{\pi a}{2}\right)\right)}{2a~\mathrm{arctanh}(a) + \ln(1 - a^2)}\right)}, \quad \text{ and } \quad g(a) = \frac{4}{\pi} - \frac{4 - \pi}{\pi}a.
\end{equation}

\section{Derivation of the surface Nusselt number}
\label{sec:surfaceNu}
It is possible to derive a surface Nusselt number by considering only the surface instead of the complete domain. The average temperatures at the interface and the bottom of the texture are $T_s = \left.\mean{T}\right|_{y = 0}$ and $T_{lt} = \left.\mean{T}\right|_{y = -k}$, respectively. We assume that $\kappa_s = \kappa_i$ and
\begin{equation}
    q = q_\mathrm{cond} + q_{\mathrm{conv},d} = \frac{\kappa_i}{k}(T_{lt} - T_s) + q_{\mathrm{conv},d}.
\end{equation}
The expression \eqref{eq:surfaceNu} for the surface Nusselt number can be derived from the definition as follows.
\begin{multline}
    \NuNumber_i = \frac{kq}{\kappa_i (T_{lt} - T_s)} = \frac{k}{\kappa_i (T_{lt} - T_s)}q_\mathrm{cond}\left(1 + \frac{q_{\mathrm{conv},d}}{q - q_{\mathrm{conv},d}}\right) \\= 1 + \frac{q_{\mathrm{conv},d}}{q - q_{\mathrm{conv},d}} = \frac{q}{q - q_{\mathrm{conv},d}} = \frac{1}{1 - q_{\mathrm{conv},d}/q}.
\end{multline}

\section{Simulations of turbulent flow}
\label{sec:numericalMethodsTurbulence}

\begin{figure}
    \centering
    \begin{subfigure}{0.45\textwidth}
        \includegraphics[width=5.5cm]{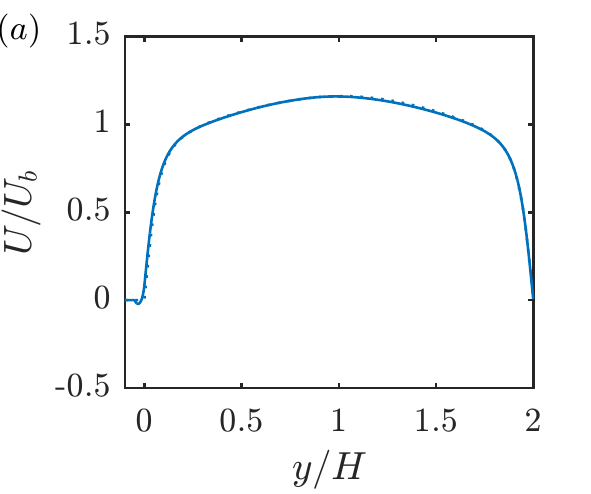}
        \captionlistentry{}
    \end{subfigure}
    \begin{subfigure}{0.45\textwidth}
        \includegraphics[width=5.5cm]{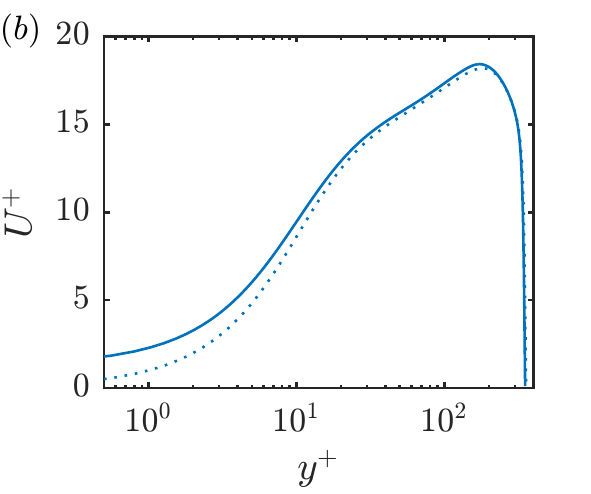}
        \captionlistentry{}
    \end{subfigure}
    \begin{subfigure}{0.45\textwidth}
        \includegraphics[width=5.5cm]{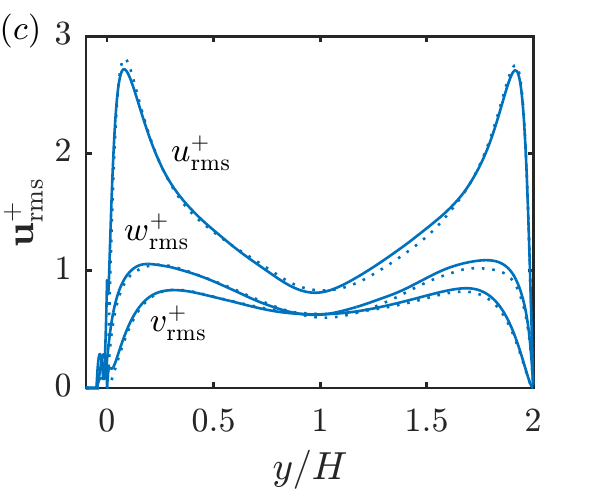}
        \captionlistentry{}
        \label{fig:turbURms}
    \end{subfigure}
    \begin{subfigure}{0.45\textwidth}
        \includegraphics[width=5.5cm]{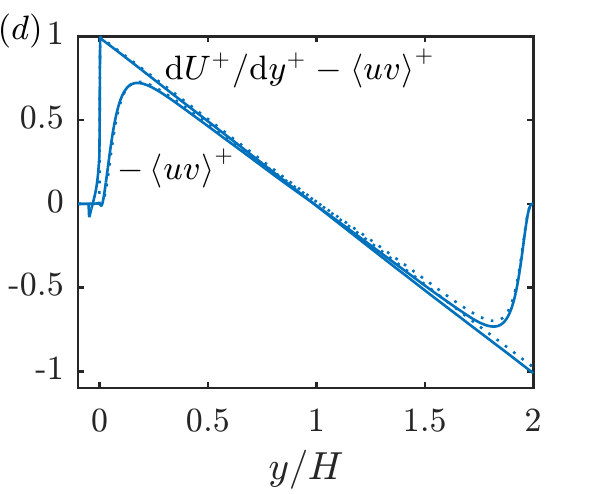}
        \captionlistentry{}
    \end{subfigure}
    \begin{subfigure}{0.45\textwidth}
        \includegraphics[width=5.5cm]{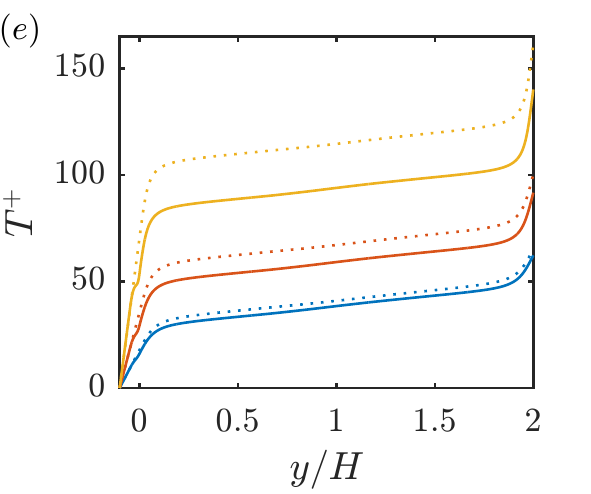}
        \captionlistentry{}
    \end{subfigure}
    \begin{subfigure}{0.45\textwidth}
        \includegraphics[width=5.5cm]{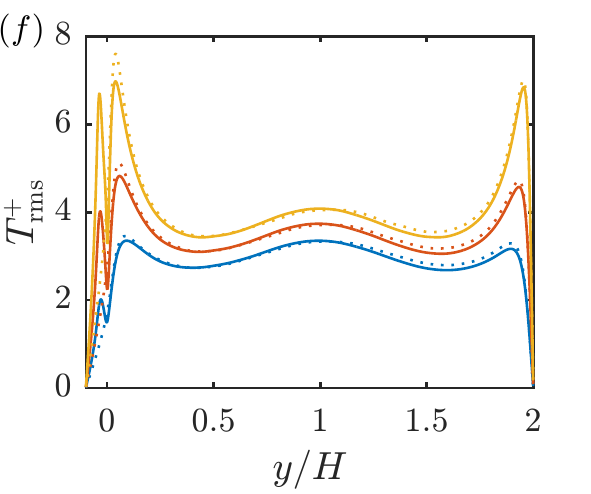}
        \captionlistentry{}
        \label{fig:turbTRms}
    \end{subfigure}
    \caption{Statistics from the turbulent flow simulations (\full, LIS; \dotted, smooth-wall reference; blue, red, and yellow for $\PrNumber_\infty = 1$, $2$, and $4$, respectively). Wall units are based on the friction velocity of the lower wall. In ($c$), $w_\mathrm{rms}^+$ denotes the spanwise root-mean-squared velocity. }
    \label{fig:turbulentStatistics}
\end{figure}

Statistics from the turbulent simulations are shown in fig.~\ref{fig:turbulentStatistics}. The mean temperature profiles are shown as $T^+ = (T_l - \mean{T})\rho c_p u_\tau /q$.

The different contributions to the surface heat flux, $q$, for the turbulent flow simulations are shown in fig.~\ref{fig:qBudgetTurb}. For these simulations, $q$ was assessed as the sum of the contributions. Deviations in the total heat flux across the channel were generally minor (around $2\%$). However, at the grooves, the errors were slightly larger. The maximum deviation occurred for $\PrNumber_\infty = 4$ ($\Delta = 0.075$, cf.~$\meanDvT^+$ in fig.~\ref{fig:turbHeatFlux} with a maximum 0.77). If $\meanDvT^+$ is assumed to have a similar error in the grooves (i.e.~$q_\mathrm{conv,d}/q \pm \Delta \cdot k/(2k + h)$), the heat transfer increase is $0.12 \le q/q^0-1 \le 0.16$ by eq.~\eqref{eq:heatFluxTurb}. Compared to the value reported in tab.~\ref{tab:turbResults}, the bounds correspond to a difference of about $2$ percentage points, which does not affect any conclusions. 

\bibliographystyle{jfm} \bibliography{references}
\end{document}